\documentclass[12pt]{article}
\usepackage{amssymb,amscd,array}
\usepackage[active]{srcltx}
\catcode `\@=11
\@addtoreset{equation}{section}

\newtheorem{thm}{Theorem}[section]

\def\qed{\blacksquare}
\newcommand{\be}{\begin{equation}}
\newcommand{\ee}{\end{equation}}
\newcommand{\bea}{\begin{eqnarray}}
\newcommand{\eea}{\end{eqnarray}}
\newcommand{\R}{\mathbb{R}}

\newcommand{\C}{\mathbb{C}}

\begin{document}
\begin{titlepage}

\begin{center}
{\bf \Large{Spin 3/2 in the Causal Approach \\}}
\end{center}
\vskip 1.0truecm
\centerline{D. R. Grigore, 
\footnote{e-mail: grigore@theory.nipne.ro}}
\vskip5mm
\centerline{Department of Theoretical Physics,}
\centerline{Institute for Physics and Nuclear Engineering ``Horia Hulubei"}
\centerline{Bucharest-M\u agurele, P. O. Box MG 6, ROM\^ANIA}

\vskip 2cm
\bigskip \nopagebreak
\vskip 1cm
\begin{abstract}
\noindent
We consider the general framework of perturbative quantum field theory for the pure Yang-Mills model developped in \cite{wick+hopf}
and consider the coupling with spin 3/2 particles. We will derive the most general form of the interaction with pure Yang-Mills particles
and with the spin $2$ field. The expressions for the interactions are obtained, as in the pure Yang-Mills case, using gauge invariance
in the first order of the perturbation theory. Further constrains are obtained using gauge invariance in the second order of the 
perturbation theory. For the coupling of spin $3/2$ with spin $1$ particles, we obtain some natural restrictions on the coupling. 
For coupling of spin $3/2$ with spin $2$ particles, we get a negative result: such a coupling is not possible in the causal approach.
\end{abstract}

\end{titlepage}

\section{Introduction}

Higher spin particles are studied in various formalisms. We mention: (a) Vasiliev formalism; (b) supersymmetry models. In this paper we 
want to study this problem in a different framework, namely the causal approach, which is an axiomatic approach based on the 
construction of the chronological products. 

We will consider only particles of spin $1, 2$ and $3/2$ (and their associated ghost fields) and determine the most general form of the 
interaction. We will prove that the interaction is unique.  

The most natural way to arrive at the Bogoliubov axioms of perturbative quantum field theory (pQFT) is by analogy with non-relativistic 
quantum mechanics \cite{Gl}, \cite{H}, \cite{D}, \cite{DF}: in this way one arrives naturally at Bogoliubov axioms 
\cite{BS}, \cite{EG}, \cite{Sc1}, \cite{Sc2}. We prefer the formulation from \cite{DF} and as presented in \cite{wick+hopf}; 
for every set of monomials 
$ 
A_{1}(x_{1}),\dots,A_{n}(x_{n}) 
$
in some jet variables (associated to some classical field theory) one associates the operator-valued distributions
$ 
T^{A_{1},\dots,A_{n}}(x_{1},\dots,x_{n})
$  
called chronological products; it will be convenient to use another notation: 
$ 
T(A_{1}(x_{1}),\dots,A_{n}(x_{n})). 
$ 

The Bogoliubov axioms express essentially some properties of the scattering matrix understood as a formal perturbation
series with the ``coefficients" the chronological products: 
(i) (skew)symmetry properties in the entries; 
(ii) Poincar\'e invariance; 
(iii) causality; 
(iv) unitarity; 
(v) the ``initial condition" which says that
$
T(A(x)) 
$
is a Wick monomial.

So we need some basic notions on free fields and Wick monomials. One can supplement these axioms by requiring: 
(vi) power counting;
(vii) Wick expansion property. 

It is a highly non-trivial problem to find solutions for the Bogoliubov axioms, even in the simplest case of a real scalar field. 
The simplest way is, in our opinion the procedure of Epstein and Glaser; it is a recursive construction for the basic objects
$ 
T(A_{1}(x_{1}),\dots,A_{n}(x_{n}))
$
and reduces the induction procedure to a distribution splitting of some distributions with causal support.  
In an equivalent way, one can reduce the induction procedure to the process of extension of distributions \cite{PS}. 

In the next Section we give some basic facts concerning Wick products, Wick submonomials, Wick theorem, pure Yang-Mills
fields and spin $2$ field (gravity in the linear approximation). In Section \ref{32} we extend the framework used for 
spins $1$ and $2$ to spin $3/2$. We also use first order gauge invariance (of the causal formalism) to determine
the most general form of the interaction between spin $3/2$ and spin $1$ and spin $3/2$ and spin $2$. In Section \ref{second}
we study gauge invariance in the  second order of the perturbation theory. In the case of the interaction between spin $3/2$ and spin $1$
we find some restrictions on the parameters of the interaction, but for the interaction between spin $3/2$ and spin $2$ we 
obtain a negative result.

\newpage
\section{Perturbative Quantum Field Theory\label{pQFT}}
There are two main ingrediants in the contruction of a perturbative quantum field theory (pQFT): the construction of the Wick monomials 
and the Bogoliubov axioms. For a pQFT of Yang-Mills theories one needs one more ingrediant, namely the introduction of ghost fields and
gauge charge; the same is true for spin $2$ particles (gravitons).

\subsection{Wick Products\label{wick prod}}

We consider a classical field theory on the Minkowski space
$
{\cal M} \simeq \R^{4}
$
(with variables
$
x^{\mu}, \mu = 0,\dots,3
$
and the metric $\eta$ with 
$
diag(\eta) = (1,-1,-1,-1)
$)
described by the Grassmann manifold 
$
\Xi_{0}
$
with variables
$
\xi_{a}, a \in {\cal A}
$
(here ${\cal A}$ is some index set) and the associated jet extension
$
J^{r}({\cal M}, \Xi_{0}),~r \geq 1
$
with variables 
$
x^{\mu},~\xi_{a;\mu_{1},\dots,\mu_{n}},~n = 0,\dots,r;
$
we denote generically by
$
\xi_{p}, p \in P
$
the variables corresponding to classical fields and their formal derivatives and by
$
\Xi_{r}
$
the linear space generated by them. The variables from
$
\Xi_{r}
$
generate the algebra
$
{\rm Alg}(\Xi_{r})
$
of polynomials.

To illustrate this, let us consider a real scalar field in Minkowski space ${\cal M}$. The first jet-bundle extension is
$$
J^{1}({\cal M}, \R) \simeq {\cal M} \times \R \times \R^{4}
$$
with coordinates 
$
(x^{\mu}, \phi, \phi_{\mu}),~\mu = 0,\dots,3.
$

If 
$
\varphi: \cal M \rightarrow \R
$
is a smooth function we can associate a new smooth function
$
j^{1}\varphi: {\cal M} \rightarrow J^{1}(\cal M, \R) 
$
according to 
$
j^{1}\varphi(x) = (x^{\mu}, \varphi(x), \partial_{\mu}\varphi(x)).
$

For higher order jet-bundle extensions we have to add new real variables
$
\phi_{\{\mu_{1},\dots,\mu_{r}\}}
$
considered completely symmetric in the indexes. For more complicated fields, one needs to add supplementary indexes to
the field i.e.
$
\phi \rightarrow \phi_{a}
$
and similarly for the derivatives. The index $a$ carries some finite dimensional representation of
$
SL(2,\C)
$
(Poincar\'e invariance) and, maybe a representation of other symmetry groups. 
In classical field theory the jet-bundle extensions
$
j^{r}\varphi(x)
$
do verify Euler-Lagrange equations. To write them we need the formal derivatives defined by
\be
d_{\nu}\phi_{\{\mu_{1},\dots,\mu_{r}\}} \equiv \phi_{\{\nu,\mu_{1},\dots,\mu_{r}\}}.
\ee

We suppose that in the algebra 
$
{\rm Alg}(\Xi_{r})
$
generated by the variables 
$
\xi_{p}
$
there is a natural conjugation
$
A \rightarrow A^{\dagger}.
$
If $A$ is some monomial in these variables, there is a canonical way to associate to $A$ a Wick 
monomial: we associate to every classical field
$
\xi_{a}, a \in {\cal A}
$
a quantum free field denoted by
$
\xi^{\rm quant}_{a}(x), a \in {\cal A}
$
and determined by the $2$-point function
\be
<\Omega, \xi^{\rm quant}_{a}(x), \xi^{\rm quant}_{b}(y) \Omega> = - i~D_{ab}^{(+)}(x - y)\times {\bf 1}.
\label{2-point}
\ee
Here 
\be
D_{ab}(x) = D_{ab}^{(+)}(x) + D_{ab}^{(-)}(x)
\ee
is the causal Pauli-Jordan distribution associated to the two fields; it is (up to some numerical factors) a polynomial
in the derivatives applied to the Pauli-Jordan distribution. We understand by 
$
D^{(\pm)}_{ab}(x)
$
the positive and negative parts of
$
D_{ab}(x)
$.
From (\ref{2-point}) we have
\be
[ \xi_{a}(x), \xi_{b}(y) ] = - i~ D_{ab}(x - y) \times {\bf 1} 
\ee
where by 
$
[\cdot, \cdot ]
$
we mean the graded commutator. 

The $n$-point functions for
$
n \geq 3
$
are obtained assuming that the truncated Wightman functions are null: see \cite{BLOT}, relations (8.74) and (8.75) and proposition 8.8
from there. The definition of these truncated Wightman functions involves the Fermi parities
$
|\xi_{p}|
$
of the fields
$
\xi_{p}, p \in P.
$

Afterwards we define
$$
\xi^{\rm quant}_{a;\mu_{1},\dots,\mu_{n}}(x) \equiv \partial_{\mu_{1}}\dots \partial_{\mu_{n}}\xi^{\rm quant}_{a}(x), a \in {\cal A}
$$
which amounts to
\be
[ \xi_{a;\mu_{1}\dots\mu_{m}}(x), \xi_{b;\nu_{1}\dots\nu_{n}}(y) ] =
(-1)^{n}~i~\partial_{\mu_{1}}\dots \partial_{\mu_{m}}\partial_{\nu_{1}}\dots \partial_{\nu_{n}}D_{ab}(x - y )\times {\bf 1}.
\label{2-point-der}
\ee
More sophisticated ways to define the free fields involve the GNS construction. 

The free quantum fields are generating a Fock space 
$
{\cal F}
$
in the sense of the Borchers algebra: formally it is generated by states of the form
$
\xi^{\rm quant}_{a_{1}}(x_{1})\dots \xi^{\rm quant}_{a_{n}}(x_{n})\Omega
$
where 
$
\Omega
$
the vacuum state.
The scalar product in this Fock space is constructed using the $n$-point distributions and we denote by
$
{\cal F}_{0} \subset {\cal F}
$
the algebraic Fock space.

One can prove that the quantum fields are free, i.e.
they verify some free field equation; in particular every field must verify Klein Gordon equation for some mass $m$
\be
(\square + m^{2})~\xi^{\rm quant}_{a}(x) = 0
\label{KG}
\ee
and it follows that in momentum space they must have the support on the hyperboloid of mass $m$. This means that 
they can be split in two parts
$
\xi^{\rm quant (\pm)}_{a}
$
with support on the upper (resp. lower) hyperboloid of mass $m$. We convene that 
$
\xi^{\rm quant (+)}_{a} 
$
resp.
$
\xi^{\rm quant (-)}_{a} 
$
correspond to the creation (resp. annihilation) part of the quantum field. The expressions
$
\xi^{\rm quant (+)}_{p} 
$
resp.
$
\xi^{\rm quant (-)}_{p} 
$
for a generic
$
\xi_{p},~ p \in P
$
are obtained in a natural way, applying partial derivatives. For a general discussion of this method of constructing free fields, 
see ref. \cite{BLOT} - especially prop. 8.8. We will follow essentially the presentation from \cite{wick+hopf}.
The Wick monomials are leaving invariant the algebraic Fock space.

\newpage
\subsection{Yang-Mills Fields\label{ym}}

First, we can generalize the preceding formalism to the case when some of the scalar fields
are odd Grassmann variables. One simply insert everywhere the Fermi sign. The next generalization is to arbitrary vector and spinorial
fields. If we consider for instance the Yang-Mills interaction Lagrangian corresponding to pure QCD then the jet variables 
$
\xi_{a}, a \in \Xi
$
are
$
(v^{\mu}_{a}, u_{a}, \tilde{u}_{a}),~a = 1,\dots,r
$
where 
$
v^{\mu}_{a}
$
are Grassmann even and 
$
u_{a}, \tilde{u}_{a}
$
are Grassmann odd variables. 

The interaction Lagrangian is determined by gauge invariance. Namely we define the {\it gauge charge} operator by
\be
d_{Q} v^{\mu}_{a} = i~d^{\mu}u_{a},\qquad
d_{Q} u_{a} = 0,\qquad
d_{Q} \tilde{u}_{a} = - i~d_{\mu}v^{\mu}_{a},~a = 1,\dots,r
\ee
where 
$
d^{\mu}
$
is the formal derivative. The gauge charge operator squares to zero:
\be
d_{Q}^{2} \simeq  0
\ee
where by
$
\simeq
$
we mean, modulo the equation of motion. Now we can define the interaction Lagrangian by the relative cohomology relation:
\be
d_{Q}T(x) \simeq {\rm total~divergence}.
\ee
If we eliminate the corresponding coboundaries, then a tri-linear Lorentz covariant 
expression is uniquely given by
\bea
T_{YM} = f_{abc} \left( {1\over 2}~v_{a\mu}~v_{b\nu}~F_{c}^{\nu\mu}
+ u_{a}~v_{b}^{\mu}~d_{\mu}\tilde{u}_{c}\right)
\label{Tint}
\eea
where
\be
F^{\mu\nu}_{a} \equiv d^{\mu}v^{\nu}_{a} - d^{\nu}v^{\mu}_{a}, 
\quad \forall a = 1,\dots,r
\ee 
and 
$
f_{abc}
$
are real and completely anti-symmetric. (This is the tri-linear part of the usual QCD interaction Lagrangian from classical field theory.)

Then we define the associated Fock space by the non-zero $2$-point distributions are
\bea
<\Omega, v^{\mu}_{a}(x_{1}) v^{\nu}_{b}(x_{2})\Omega> = 
i~\eta^{\mu\nu}~\delta_{ab}~D_{0}^{(+)}(x_{1} - x_{2}),
\nonumber \\
<\Omega, u_{a}(x_{1}) \tilde{u}_{b}(x_{2})\Omega> = - i~\delta_{ab}~D_{0}^{(+)}(x_{1} - x_{2}),
\nonumber\\
<\Omega, \tilde{u}_{a}(x_{1}) u_{b}(x_{2})\Omega> = i~\delta_{ab}~D_{0}^{(+)}(x_{1} - x_{2}).
\label{2-massless-vector}
\eea
and we have the causal commutation relations 
\be
~[v_{\mu}(x_{1}), v_{\mu}(x_{2}) ] =i~\eta_{\mu\nu}~D_{0}(x_{1} - x_{2})~\cdot I,
\qquad
[u(x_{1}), \tilde{u}(x_{2})] = - i~D_{0}(x_{1} - x_{2})~\cdot I
\ee
and the other commutators are null. In (\ref{2-massless-vector}) we are using the Pauli-Jordan distribution
\be
D_{m}(x) = D_{m}^{(+)}(x) + D_{m}^{(-)}(x)
\ee
where
\be
D_{m}^{(\pm)}(x) =
\pm {i \over (2\pi)^{3}}~\int dp e^{- i p\cdot x} \theta(\pm p_{0}) \delta(p^{2} -m^{2}).
\ee

Then we construct the associated Wick monomials. Then the expression (\ref{Tint}) gives a Wick polynomial 
$
T^{\rm quant}
$
formally the same, but: 
(a) the jet variables must be replaced by the associated quantum fields; (b) the formal derivative 
$
d^{\mu}
$
goes in the true derivative in the coordinate space; (c) Wick ordering should be done to obtain well-defined operators. We also 
have an associated {\it gauge charge} operator in the Fock space given by
\bea
~[Q, v^{\mu}_{a}] = i~\partial^{\mu}u_{a},\qquad
\{ Q, u_{a} \} = 0,\qquad
\{Q, \tilde{u}_{a}\} = - i~\partial_{\mu}v^{\mu}_{a}
\nonumber \\
Q \Omega = 0.
\label{Q-vector-null}
\eea

Then it can be proved that
$
Q^{2} = 0
$
and is well defined i.e. it should preserve the canonical commutation relation. In our case this reduces to
\be
[Q, [ v_{\mu}(x_{1}),\tilde{u}(x_{2})]] + {\rm cyclic~permutations} = 0.
\ee

It is important point that the preceding construction leads to a true Hilbert space: the sesqui-linear form 
$
<\cdot,\cdot>
$
is positively defined on
$
Ker(Q)/Im(Q).
$

We recall briefly the argument from \cite{cohomology}. The generic form of an one-particle state is
\be
\Psi = \left[ \int f_{\mu}(x) v^{\mu}(x) + \int g_{1}(x) u(x) + \int g_{2}(x) \tilde{u}(x) \right] \Omega
\ee
with test functions
$
f_{\mu}, g_{1}, g_{2}
$
verifying the wave equation equation. We impose the condition 
$
\Psi \in Ker(Q) \quad \Longleftrightarrow \quad Q\Psi = 0;
$
we obtain 
$
\partial^{\mu}f_{\mu} = 0
$
and
$
g_{2} = 0
$
i.e. the generic element
$
\Psi \in Ker(Q)
$
is
\be
\Psi = \left[ \int f_{\mu}(x) v^{\mu}(x) + \int g(x) u(x) \right] \Omega
\label{kerQ-0}
\ee
with $g$ arbitrary and 
$
f_{\mu}
$
constrained by the transversality condition 
$
\partial^{\mu}f_{\mu} = 0;
$
so the elements of
$
Ker(Q)
$
are in one-one correspondence with couples of test functions
$
(f_{\mu}, g)
$
with the transversality condition on the first entry. Now, a generic element
$
\Psi^{\prime} \in Ran(Q)
$
has the form 
\be
\Psi^{\prime} = Q\Phi = \left[\int \partial_{\mu}g^{\prime}(x) v^{\mu}(x) 
- \int \partial^{\mu}f^{\prime}_{\mu}(x) u(x) \right] \Omega
\label{ranQ-0}
\ee
so if
$
\Psi \in Ker(Q)
$
is indexed by the couple 
$
(f_{\mu}, g)
$
then 
$
\Psi + \Psi^{\prime}
$
is indexed by the couple
$
(f_{\mu} + \partial_{\mu}g^{\prime}, g - \partial^{\mu}f^{\prime}_{\mu}).
$
If we take 
$
f^{\prime}_{\mu}
$
conveniently we can make 
$
g = 0.
$
We introduce the equivalence relation 
$
f_{\mu}^{(1)} \sim f_{\mu}^{(2)} \quad \Longleftrightarrow 
f_{\mu}^{(1)} - f_{\mu}^{(2)} = \partial_{\mu}g^{\prime}
$
and it follows that the equivalence classes are indexed by equivalence classes of wave functions
$
[f_{\mu}];
$
it remains to prove that the sesquilinear form 
$<\cdot,\cdot>$ 
induces a positively defined form on
$
Ker(Q))/ Ran(Q))
$ 
and we have obtained the usual one-particle Hilbert space for the photon. In momentum space the transversality condition
is:
\be
p^{\mu}~\tilde{f}_{\mu} = 0
\ee
where $\tilde{f}$ is the Fourier transform of $f$. In the frame where
$
p = (1,0,0,1)
$
we obtain
$
\tilde{f}^{0} - \tilde{f}^{3} = 0
$
so
$
\tilde{f}^{\mu} = ( \tilde{f}^{0}, \tilde{f}^{1}, \tilde{f}^{2}, \tilde{f}^{0}) \sim  (0, \tilde{f}^{1}, \tilde{f}^{2}, 0) 
$
so
$
< f, f> \sim |\tilde{f}^{1}|^{2} + |\tilde{f}^{2}|^{2} \geq 0.
$

The preceding procedure leads to
\be
~[Q, T^{\rm quant}(x) ] = {\rm total~divergence}
\label{gauge1}
\ee
where the equations of motion are automatically used because the quantum fields are on-shell.
From now on we abandon the super-script {\it quant} because it will be obvious from the context if we refer 
to the classical expression (\ref{Tint}) or to its quantum counterpart.

We conclude our presentation with a generalization of (\ref{gauge1}). In fact, it can be proved that (\ref{gauge1}) implies
the existence of Wick polynomials
$
T^{\mu}
$
and
$
T^{\mu\nu}
$
such that we have:
\be
~[Q, T^{I} ] = i \partial_{\mu}T^{I\mu}
\label{gauge2}
\ee
for any multi-index $I$ with the convention
$
T^{\emptyset} \equiv T.
$
Explicitly:
\bea
T^{\mu} = f_{abc} \left( u_{a}~v_{b\nu}~F_{c}^{\nu\mu}
- {1\over 2}u_{a}~u_{b}~d^{\mu}\tilde{u}_{c}\right)
\label{Tmu-int}
\eea
and 
\bea
T^{\mu\nu} = {1\over 2}~f_{abc}~u_{a}~u_{b}~F_{c}^{\mu\nu}.
\label{Tmunu-int}
\eea

We now give the relation expressing gauge invariance in order $n$ of the perturbation theory. We define the operator 
$
\delta
$
on chronological products by:
\bea
\delta T(T^{I_{1}}(x_{1}),\dots,T^{I_{n}}(x_{n})) \equiv 
\sum_{m=1}^{n}~( -1)^{s_{m}}\partial_{\mu}^{m}T(T^{I_{1}}(x_{1}), \dots,T^{I_{m}\mu}(x_{m}),\dots,T^{I_{n}}(x_{n}))
\label{derT}
\eea
with
\be
s_{m} \equiv \sum_{p=1}^{m-1} |I_{p}|,
\ee
then we define the operator
\be
s \equiv d_{Q} - i \delta.
\label{s-n}
\ee

Gauge invariance in an arbitrary order is then expressed by
\be
sT(T^{I_{1}}(x_{1}),\dots,T^{I_{n}}(x_{n})) = 0.
\label{brst-n}
\ee

If we impose gauge invariance in the second order of the perturbation theory then we obtain that the constants
$
f_{abc}
$
must verify Jacobi identity. The proof of this assertion can be found in \cite{wick+hopf} and is based on the use of the 
Wick submonomials
\bea
B_{a\mu} \equiv \tilde{u}_{a,\mu} \cdot T = - f_{abc} u_{b}~v_{c\mu}
\nonumber\\
C_{a\mu} \equiv v_{a\mu} \cdot T = f_{abc} (v_{b}^{\nu}~F_{c\nu\mu} - u_{b}~\tilde{u}_{c,\mu})
\nonumber\\
D_{a} \equiv u_{a} \cdot T = f_{abc} v^{\mu}_{b}~\tilde{u}_{c,\mu}
\nonumber\\
E_{a\mu\nu} \equiv v_{a\mu,\nu} \cdot T = f_{abc} v_{b\mu}~v_{c\nu}
\nonumber\\
C_{a\nu\mu} \equiv v_{a\nu} \cdot T_{\mu} = - f_{abc} u_{b}~F_{c\nu\mu}.
\label{sub1}
\eea
We also have
\bea
u_{a} \cdot T = - C_{a\mu}
\nonumber\\
v_{a\rho,\sigma} \cdot T_{\mu} = \eta_{\mu\sigma}~B_{a\rho} -  \eta_{\mu\rho}~B_{a\sigma}
\nonumber\\
u_{a} \cdot T_{\mu\nu} = - C_{a\mu\nu}
\label{sub2}
\eea
If we define
\be
B_{a} \equiv {1 \over 2}~f_{abc}~u_{b}~u_{c}
\label{B}
\ee
we also have
\bea
\tilde{u}_{a,\nu} \cdot T_{\mu} = \eta_{\mu\nu}~B_{a}
\nonumber\\
v_{a\rho,\sigma} \cdot T_{\mu\nu} = (\eta_{\mu\sigma}~\eta_{\nu\rho} - \eta_{\nu\sigma}~\eta_{\mu\rho})~B_{a}.
\label{sub3}
\eea
We have proved in \cite{wick+hopf} that
\be
s^{\prime} A = 0
\ee
where 
\be
s \equiv d_{Q} - i\delta, \qquad s^{\prime} \equiv s - i\delta^{\prime} = d_{Q} - i(\delta +\delta^{\prime}).
\ee
and
\be
\delta^{\prime}E_{a}^{\mu\nu} = C_{a}^{\mu\nu}.
\label{dprime}
\ee
and $0$ for the other Wick submonomials.
\newpage
\subsection{Quantum Gravity in the Linear Approximation\label{gr}}

If we consider pure massless gravity then the jet variables are 
$
(h_{\mu\nu}, u_{\rho}, \tilde{u}_{\sigma})
$
where 
$
h_{\mu\nu}
$
are Grassmann even and  
$
u_{\rho}, \tilde{u}_{\sigma}
$
are Grassmann odd variables. 
The interaction Lagrangian is determined by gauge invariance. Namely we define the {\it gauge charge} operator by
\bea
d_{Q} h_{\mu\nu} = -{i\over 2}~\Bigl( d_{\mu}u_{\nu} + d_{\nu}u_{\mu} - \eta_{\mu\nu}~d_{\rho}u^{\rho}\Bigl),
\nonumber\\
d_{Q} u_{\rho} = 0,
\qquad
d_{Q} \tilde{u}_{\sigma} = i~d^{\lambda}h_{\sigma\lambda}.
\label{Q-h}
\eea
The gauge charge operator squares to zero:
\be
d_{Q}^{2} \simeq  0
\ee
where by
$
\simeq
$
we mean, modulo the equation of motion. 

Then we define the associated Fock space by the non-zero $2$-point distributions are
\bea
<\Omega, h_{\mu\nu}(x_{1}) h_{\rho\sigma}(x_{2})\Omega> = 
- {i\over 2}~\Bigl(\eta_{\mu\rho}~\eta_{\nu\sigma} + \eta_{\mu\sigma}~\eta_{\nu\rho} 
- \eta_{\mu\nu}\eta_{\rho\sigma}\Bigl)~D_{0}^{(+)}(x_{1} - x_{2})
\nonumber \\
<\Omega, u_{\rho}(x_{1}) \tilde{u}_{\sigma}(x_{2})\Omega> = i~\eta_{\rho\sigma}~D_{0}^{(+)}(x_{1} - x_{2}),
\nonumber\\
<\Omega, \tilde{u}_{\sigma}(x_{1}) u_{\rho}(x_{2})\Omega> = - i~\eta_{\rho\sigma}~D_{0}^{(+)}(x_{1} - x_{2}).
\label{2-h}
\eea

We will use the notation
\be
h \equiv \eta^{\mu\nu}~h_{\mu\nu}.
\ee

The definitions above are describing a system of massless particles of helicity $2$ as it is proved in \cite{gravity}. 
It can be proved \cite{ren-gravity} that one must take  
\be
h_{\mu\nu}\cdot h_{\rho\sigma} \equiv {1\over 2}~(\eta_{\mu\rho}~\eta_{\nu\sigma} + \eta_{\mu\sigma}~\eta_{\nu\rho}).
\ee
\newpage

\section{Particles of spin 3/2\label{32}}

We generalize the framework of the preceding Section to particles of spin $3/2$. The jet variables 
$
\xi_{a}, a \in \Xi
$
are in this case
$
(\Psi_{\mu}, \chi, \tilde{\chi})
$
where 
$
\Psi_{\mu} = (\Psi_{\mu\alpha})_{\alpha = 1,\dots,4}
$
is a spinor Fermi field (Grassmann odd, and of ghost number $0$) and 
$
\chi, \tilde{\chi}$
are  the associated ghost fields (Grassmann even and of ghost number $1$, resp. $- 1$). 

The interaction Lagrangian is determined by gauge invariance. Namely we define the {\it gauge charge} operator by
\be
d_{Q} \Psi_{\mu} = i~d^{\mu}\chi,\qquad
d_{Q} \chi = 0,\qquad
d_{Q} \tilde{\chi} = i~d^{\mu}\Psi_{\mu}
\ee
where 
$
d^{\mu}
$
is the formal derivative. The gauge charge operator squares to zero:
\be
d_{Q}^{2} \simeq  0
\ee
where by
$
\simeq
$
we mean, modulo the equation of motion. These equations are in this case the Dirac equations for null mass:
\be
\gamma\cdot d~\Psi_{\mu} = 0, \quad  \gamma\cdot~d~\chi = 0, \quad \gamma\cdot~d~\tilde{\chi} = 0
\ee
where
$
\gamma_{\rho}
$
are the Dirac matrices and
$
\gamma\cdot d \equiv \gamma_{\lambda} d^{\lambda}.
$

Now we can define the interaction Lagrangian, as in the pure Yang-Mills case, by the relative cohomology (or gauge invariance) 
relation:
\be
d_{Q}T \sim {\rm total~divergence}.
\ee

We will prove that this condition determines uniquely the expression $T$ up to coboundaries. 
\begin{thm}
We consider the polynomials $T$ and $T^{\mu}$ depending on the pure Yang-Mills variables
$
(v^{\mu}_{a}, u_{a}, \tilde{u}_{a}),~a = 1,\dots,r
$
and on
$
(\Psi_{\mu}, \chi, \tilde{\chi})
$
such that the dependence on 
$
(\Psi_{\mu}, \chi, \tilde{\chi})
$
is non-trivial. Also we impose

(i) the polynomials are Lorentz covariant and do verify
$
\omega(T), \omega(T^{\mu}) \leq 4
$
and
$
gh(T) = 0, gh(T^{\mu}) = 1;
$

(ii) the gauge invariance condition
\be
sT \equiv d_{Q}T - i~d_{\mu}T^{\mu} = 0.
\label{gauge1a}
\ee
is true.

Then the expression $T$ is, up to a coboundary
\be
T = v_{a}^{\mu}~\bar{\Psi}^{\mu}~t_{a}^{\epsilon} \otimes \gamma_{\mu}\gamma_{\rho}\gamma_{\nu}\gamma_{\epsilon}\Psi^{\nu}
+ 2~u_{a}~(\bar{\Psi}^{\mu}~t_{a}^{\epsilon} \otimes \gamma_{\mu}\gamma_{\epsilon}\tilde{\chi}
- \bar{\tilde{\chi}} t_{a}^{\epsilon} \otimes \gamma_{\mu}\gamma_{\epsilon}\Psi^{\mu})
\ee
and we also have
\be
T^{\mu} = u_{a}~\bar{\Psi}_{\nu}~t_{a}^{\epsilon} \otimes \gamma^{\nu}\gamma^{\mu}\gamma^{\rho}\gamma_{\epsilon}\Psi_{\rho}
\ee
\be
d_{Q}T^{\mu} = 0.
\ee
Here we have used the notation
\be
\gamma_{\epsilon} \equiv {1\over 2} (1 + \epsilon \gamma_{5}), \qquad 
\gamma_{5} \equiv i~\gamma_{0}\gamma_{1}\gamma_{2}\gamma_{3}
\ee
and
\be
\bar{\Psi} \equiv \Psi^{+}~\gamma_{0}, \quad \bar{\chi} \equiv \chi^{+}~\gamma_{0}, \quad
\bar{\tilde{\chi}} \equiv \tilde{\chi}^{+}~\gamma_{0}.
\ee
The expression $T$ is self-adjoint if the matrices
$
t_{a}^{\epsilon}
$
are self-adjoint.
\label{3/2+1}
\end{thm}
{\bf Proof:}
The generic expression for $T$ depending non-trivially on the variables
$
(\Psi_{\mu}, \chi, \tilde{\chi})
$
is a summ of the following contributions:
\bea
T_{1} = v_{a}^{\rho}~\bar{\Psi}_{\mu}~t_{a\epsilon}^{(1)} \otimes \gamma_{\rho}\gamma_{\epsilon}\Psi^{\mu}, \qquad
T_{2} = v_{a}^{\rho}~\bar{\Psi}_{\rho}~t_{a\epsilon}^{(2)} \otimes \gamma_{\mu}\gamma_{\epsilon}\Psi^{\mu}
\nonumber\\
T_{3} = v_{a}^{\rho}~\bar{\Psi}^{\mu}~t_{a\epsilon}^{(3)} \otimes \gamma_{\mu}\gamma_{\epsilon}\Psi_{\rho}, \qquad
T_{4} = v_{a}^{\rho}~
\bar{\Psi}^{\mu}~t_{a\epsilon}^{(4)} \otimes \gamma_{\mu}\gamma_{\rho}\gamma_{\nu}\gamma_{\epsilon}\Psi^{\nu}
\nonumber\\
T_{5} = v_{a}^{\rho}~\bar{\tilde{\chi}}~t_{a\epsilon}^{(5)} \otimes \gamma_{\rho}\gamma_{\epsilon}\chi, \qquad
T_{6} = v_{a}^{\rho}~\bar{\chi}~t_{a\epsilon}^{(6)} \otimes \gamma_{\mu}\gamma_{\epsilon}\tilde{\chi}
\nonumber\\
T_{7} = u_{a}~\bar{\tilde{\chi}}~t_{a\epsilon}^{(7)} \otimes \gamma_{\mu}\gamma_{\epsilon}\Psi^{\mu}, \qquad
T_{8} = u_{a}~\bar{\Psi}_{\mu}~t_{a\epsilon}^{(8)} \otimes \gamma^{\mu}\gamma_{\epsilon}\tilde{\chi}
\nonumber\\
T_{9} = \tilde{u}_{a}~\bar{\Psi}_{\mu}~t_{a\epsilon}^{(9)} \otimes \gamma^{\mu}\gamma_{\epsilon}\chi, \qquad
T_{10} = \tilde{u}_{a}~\bar{\chi}~t_{a\epsilon}^{(10)} \otimes \gamma^{\mu}\gamma_{\epsilon}\Psi_{\mu}.
\eea

In the same way, the generic form of 
$
T^{\mu}
$
is a sum of the following terms:
\bea
T_{1}^{\mu} = v_{a}^{\rho}~\bar{\chi}~s_{a\epsilon}^{(1)} \otimes \gamma_{\rho}\gamma_{\epsilon}\Psi^{\mu}, \qquad
T_{2}^{\mu} = v_{a}^{\rho}~\bar{\chi}~s_{a\epsilon}^{(2)} \otimes \gamma^{\mu}\gamma_{\epsilon}\Psi^{\rho}
\nonumber\\
T_{3}^{\mu} = v_{a}^{\rho}~\bar{\Psi}^{\mu}~s_{a\epsilon}^{(3)} \otimes \gamma_{\rho}\gamma_{\epsilon}\chi, \qquad
T_{4}^{\mu} = v_{a}^{\rho}~\bar{\Psi}_{\rho}~s_{a\epsilon}^{(4)} \otimes \gamma^{\mu}\gamma_{\epsilon}\chi
\nonumber\\
T_{5}^{\mu} = v_{a}^{\mu}~\bar{\chi}~s_{a\epsilon}^{(5)} \otimes \gamma_{\rho}\gamma_{\epsilon}\Psi^{\rho}, \qquad
T_{6}^{\mu} = v_{a}^{\mu}~\bar{\Psi}_{\rho}~s_{a\epsilon}^{(6)} \otimes \gamma^{\rho}\gamma_{\epsilon}\chi
\nonumber\\
T_{7}^{\mu} = v_{a}^{\rho}~\bar{\chi}~s_{a\epsilon}^{(7)} 
\otimes \gamma^{\mu}\gamma_{\rho}\gamma_{\nu}\gamma_{\epsilon}\Psi^{\nu}, \qquad
T_{8}^{\mu} = v_{a}^{\rho}~\bar{\Psi}^{\nu}~s_{a\epsilon}^{(8)} 
\otimes \gamma^{\mu}\gamma_{\rho}\gamma_{\nu}\gamma_{\epsilon}\chi
\nonumber\\
T_{9}^{\mu} = u_{a}~\bar{\Psi}_{\rho}~s_{a\epsilon}^{(9)} \otimes \gamma^{\mu}\gamma_{\epsilon}\Psi^{\rho}, \qquad
T_{10}^{\mu} = u_{a}~\bar{\Psi}^{\mu}~s_{a\epsilon}^{(10)} \otimes \gamma_{\rho}\gamma_{\epsilon}\Psi_{\rho}
\nonumber\\
T_{11}^{\mu} = u_{a}~\bar{\Psi}_{\rho}~s_{a\epsilon}^{(11)} \otimes \gamma^{\rho}\gamma_{\epsilon}\Psi^{\mu}, \qquad
T_{12}^{\mu} = u_{a}~\bar{\tilde{\chi}}~s_{a\epsilon}^{(12)} \otimes \gamma^{\mu}\gamma_{\epsilon}\chi
\nonumber\\
T_{13}^{\mu} = u_{a}~\bar{\chi}~s_{a\epsilon}^{(13)} \otimes \gamma^{\mu}\gamma_{\epsilon}\tilde{\chi}, \qquad
T_{14}^{\mu} = u_{a}~\bar{\Psi}_{\nu}~s_{a\epsilon}^{(14)} \otimes 
\gamma^{\nu}\gamma^{\mu}\gamma^{\rho}\gamma_{\epsilon}\Psi_{\rho}
\nonumber\\
T_{15} = \tilde{u}_{a}~\bar{\chi}~s_{a\epsilon}^{(15)} \otimes \gamma^{\mu}\gamma_{\epsilon}\chi.
\eea

If we substitute in (\ref{gauge1a}) we find out easily the solution
$
s_{14} = t_{4}, \quad t_{7} = - 2~ t_{4}, \quad t_{8} = 2~ t_{4}
$
which is the solution from the statment with
$
t \equiv t_{4}.
$

For the self-adjoncy problem we use the identities
\bea
(\bar{\Psi}_{1}~\gamma_{\mu}\gamma_{\epsilon}\Psi_{2})^{*}
= \bar{\Psi}_{2}~\gamma_{\mu}\gamma_{\epsilon}\Psi_{2}
\nonumber\\
(\bar{\Psi}_{1}~\gamma_{\mu}\gamma_{\nu}\gamma_{\rho}\gamma_{\epsilon}\Psi_{2})^{*}
= \bar{\Psi}_{2}~\gamma_{\rho}\gamma_{\nu}\gamma_{\mu}\gamma_{\epsilon}\Psi_{2}
\eea
which follows from
\be
\gamma_{0}\gamma_{\mu}^{*}\gamma_{0} = \gamma_{\mu}
\ee
and are valid regardless of the Grassmann nature of the fields
$
\Psi_{1}, \Psi_{2}.
$
$\qed$

To construct the associated Fock space we define the non-zero $2$-point distributions by:
\bea
<\Omega, \Psi_{\mu\alpha}(x_{1}) \bar{\Psi}_{\nu\beta}(x_{2})\Omega> = 
i~\eta^{\mu\nu}~S_{0}^{(+)}(x_{1} - x_{2})_{\alpha\beta},
\nonumber\\
<\Omega, \bar{\Psi}_{\mu\alpha}(x_{1}) \Psi_{\nu\beta}(x_{2})\Omega> = 
i~\eta^{\mu\nu}~S_{0}^{(-)}(x_{1} - x_{2})_{\beta\alpha},
\nonumber \\
<\Omega, \chi_{\alpha}(x_{1}) \bar{\tilde{\chi}}_{\beta}(x_{2})\Omega> = - i~S_{0}^{(+)}(x_{1} - x_{2})_{\alpha\beta},
\nonumber\\
<\Omega, \bar{\tilde{\chi}}_{\alpha}(x_{1}) \chi_{\beta}(x_{2})\Omega> = i~S_{0}^{(+)}(x_{1} - x_{2})_{\beta\alpha}.
\label{2-massless-3/2}
\eea
where we are using 
\be
S^{(\pm)}(x) = i\gamma\cdot\partial D^{(\pm)}
\ee
and construct the associated Wick monomials. One obtains from these relations the canonical commutation relations:
\bea
~[ \Psi_{\mu\alpha}(x_{1}), \bar{\Psi}_{\nu\beta}(x_{2}) ] = i~\eta^{\mu\nu}~S_{0}(x_{1} - x_{2})_{\alpha\beta},
\nonumber\\
~[ \chi_{\alpha}(x_{1}), \bar{\tilde{\chi}}_{\beta}(x_{2}) ] = - i~\eta^{\mu\nu}~S_{0}(x_{1} - x_{2})_{\alpha\beta},
\nonumber \\
~[ \tilde{\chi}_{\alpha}(x_{1}), \bar{\chi}_{\beta}(x_{2}) ] = - i~\eta^{\mu\nu}~S_{0}(x_{1} - x_{2})_{\alpha\beta}.
\eea
Then the expression (\ref{Tint}) gives a Wick polynomial 
$
T^{\rm quant}
$
formally the same, but: 
(a) the jet variables must be replaced by the associated quantum fields; (b) the formal derivative 
$
d^{\mu}
$
goes in the true derivative in the coordinate space; (c) Wick ordering should be done to obtain well-defined operators. We also 
have an associated {\it gauge charge} operator in the Fock space given by
\bea
\{Q, \Psi_{\mu}\} = i~\partial_{\mu}\chi,\quad \Leftrightarrow \quad \{Q, \bar{\Psi}_{\mu}\} = - i~\partial_{\mu}\bar{\chi},
\nonumber\\
~[ Q, \chi ] = 0,\quad \Leftrightarrow \quad [ Q, \bar{\chi} ] = 0,
\nonumber\\
~[ Q, \tilde{\chi} ] = i~\partial^{\mu}\Psi_{\mu}, \quad \Leftrightarrow \quad 
[ Q, \bar{\tilde{\chi}} ] = i~\partial^{\mu}\bar{\Psi}_{\mu}
\nonumber \\
Q \Omega = 0.
\label{Q-3/2}
\eea

Then it can be proved that
$
Q^{2} = 0
$
and $Q$ is well defined i.e. it should preserve the canonical commutation relation. In our case this reduces to
\be
[Q, [ \Psi_{\mu\alpha}(x_{1}),\tilde{\chi}_{\beta}(x_{2})]] + {\rm cyclic~permutations} = 0
\ee
where 
$
[\cdot,\cdot ]
$
is the graded commutator. The preceding construction leads to a true Hilbert space: the sesqui-linear form 
$
<\cdot,\cdot>
$
is positively defined on
$
Ker(Q)/Im(Q)
$
the argument being similar to the argument from the Yang-Mills case, but technically more involved because of the 
spinorial structure.

Moreover we have (\ref{gauge1}):

\be
~[Q, T^{\rm quant}(x) ] = {\rm total~divergence}
\label{gauge1-3/2}
\ee
where the equations of motion are automatically used because the quantum fields are on-shell.
From now on we abandon the super-script {\it quant} because it will be obvious from the context if we refer 
to the classical expression (\ref{Tint}) or to its quantum counterpart.
\newpage
In the same way one can analyse the interaction between the field of spin $3/2$ and (massless) gravity. 
\begin{thm}
We consider the polynomials $T$ and $T^{\mu}$ depending on the gravity variables
$
(h_{\mu\nu}, u_{\rho}, \tilde{u}_{\sigma})
$
and on a set of spin $3/2$ fields
$
(\Psi^{\mu}_{A}, \chi_{A}, \tilde{\chi}_{A}), A = 1,\dots,N
$
such that the dependence on 
$
(\Psi_{\mu}, \chi, \tilde{\chi})
$
is non-trivial. Also we impose

(i) the polynomials are Lorentz covariant and do verify
$
\omega(T), \omega(T^{\mu}) \leq 5
$
and
$
gh(T) = 0, gh(T^{\mu}) = 1;
$

(ii) the gauge invariance condition (\ref{gauge1a}) is true.

Then the expression $T$ is, up to a coboundary, the sum of the following two expressions
\bea
t_{1} = i~h_{\mu\nu}~( \bar{\Psi}^{\rho}~t_{1}^{\epsilon} \otimes 
\gamma_{\rho}\gamma^{\mu}\gamma_{\sigma}\gamma_{\epsilon}d^{\nu}\Psi^{\sigma} 
- d^{\nu}\bar{\Psi}^{\sigma}~t_{1}^{\epsilon} \otimes 
\gamma_{\sigma}\gamma^{\mu}\gamma_{\rho}\gamma_{\epsilon}\Psi^{\rho} )
\nonumber\\
+ 2 i~u_{\mu}~( d^{\mu}\bar{\Psi}^{\nu}~t_{1}^{\epsilon} \otimes \gamma_{\nu}\gamma_{\epsilon}\tilde{\chi}
+ \bar{\tilde{\chi}} t_{1}^{\epsilon} \otimes \gamma_{\nu}\gamma_{\epsilon}d^{\mu}\Psi^{\nu})
\eea
and
\bea
t_{2} = i~h_{\mu\nu}~( \bar{\Psi}_{\rho}~t_{2}^{\epsilon} \otimes \gamma^{\mu}\gamma_{\epsilon}d^{\nu}\Psi^{\sigma} 
- d^{\nu}\bar{\Psi}_{\rho}~t_{2}^{\epsilon} \otimes \gamma^{\mu}\gamma_{\epsilon}\Psi^{\rho} )
\nonumber\\
+ 2 i~h_{\mu\nu}~( \bar{\Psi}_{\rho}~t_{2}^{\epsilon} \otimes \gamma^{\mu}\gamma_{\epsilon}d^{\rho}\Psi^{\nu} 
- d^{\rho}\bar{\Psi}^{\nu}~t_{2}^{\epsilon} \otimes \gamma^{\mu}\gamma_{\epsilon}\Psi_{\rho} )
\nonumber\\
- i~h~( \bar{\Psi}^{\mu}~t_{2}^{\epsilon} \otimes \gamma^{\nu}\gamma_{\epsilon}d_{\mu}\Psi_{\nu} 
- d_{\mu}\bar{\Psi}_{\nu}~t_{2}^{\epsilon} \otimes \gamma^{\nu}\gamma_{\epsilon}\Psi^{\mu} )
\nonumber\\
- i~u_{\mu}~( \bar{\Psi}_{\nu}~t_{2}^{\epsilon} \otimes \gamma^{\mu}\gamma_{\epsilon}d^{\nu}\tilde{\chi}
+ d^{\nu}\bar{\tilde{\chi}} t_{2}^{\epsilon} \otimes \gamma^{\mu}\gamma_{\epsilon}\Psi_{\nu})
\nonumber\\
- i~\tilde{u}_{\mu}~( \bar{\Psi}_{\nu}~t_{2}^{\epsilon} \otimes \gamma^{\mu}\gamma_{\epsilon}d^{\nu}\chi
- d^{\nu}\bar{\chi} t_{2}^{\epsilon} \otimes \gamma^{\mu}\gamma_{\epsilon}\Psi_{\nu})
\eea

The expression $T$ is self-adjoint if the matrices
$
t_{a}^{\epsilon}, a = 1,2
$
are self-adjoint.

We also have
\be
d_{Q} t_{j} \sim i d_{\mu}t_{j}^{\mu}, \quad j = 1,2
\ee
where
\bea
t_{1}^{\mu} = i~u^{\mu}~( \bar{\Psi}^{\nu} t_{1}^{\epsilon} \otimes \gamma_{\nu} \gamma_{\epsilon} d_{\rho}\Psi^{\rho}
- d_{\nu}\bar{\Psi}^{\nu} t_{1}^{\epsilon} \otimes \gamma_{\rho} \gamma_{\epsilon} \Psi^{\rho})
\nonumber\\
- {i \over 2}~u_{\nu}~(\bar{\Psi}^{\rho} t_{1}^{\epsilon} \otimes 
\gamma_{\rho}\gamma^{\nu}\gamma_{\sigma} \gamma_{\epsilon} d^{\mu}\Psi^{\sigma}
- d^{\mu}\bar{\Psi}^{\rho} t_{1}^{\epsilon} \otimes 
\gamma_{\rho}\gamma^{\nu}\gamma_{\sigma} \gamma_{\epsilon} \Psi^{\sigma})
\nonumber\\
- {i \over 2}~u_{\nu}~(\bar{\Psi}^{\rho} t_{1}^{\epsilon} \otimes 
\gamma_{\rho}\gamma^{\mu}\gamma_{\sigma} \gamma_{\epsilon} d^{\nu}\Psi^{\sigma}
- d^{\nu}\bar{\Psi}^{\rho} t_{1}^{\epsilon} \otimes 
\gamma_{\rho}\gamma^{\mu}\gamma_{\sigma} \gamma_{\epsilon} \Psi^{\sigma})
\eea
\be
t_{1}^{\mu\nu} = 0, \qquad t_{1}^{\mu\nu\rho} = 0
\ee
and
\bea
t_{2}^{\mu} = {i \over 2}~u^{\nu}~(\bar{\Psi}^{\rho} t_{2}^{\epsilon} \otimes \gamma^{\mu} \gamma_{\epsilon} d_{\nu}\Psi_{\rho}
- d_{\nu}\bar{\Psi}_{\rho} t_{2}^{\epsilon} \otimes \gamma^{\mu} \gamma_{\epsilon} \Psi^{\rho})
\nonumber\\
+ i~(- \bar{\Psi}^{\rho} t_{2}^{\epsilon} \otimes \gamma^{\mu} \gamma_{\epsilon} d_{\rho}\Psi_{\nu}
+ d_{\rho}\bar{\Psi}_{\nu} t_{2}^{\epsilon} \otimes \gamma^{\mu} \gamma_{\epsilon} \Psi^{\rho})
+ {i \over 2}~u^{\nu}~(\bar{\Psi}_{\rho} t_{2}^{\epsilon} \otimes \gamma_{\nu} \gamma_{\epsilon} d^{\mu}\Psi^{\rho}
- d^{\mu}\bar{\Psi}_{\rho} t_{2}^{\epsilon} \otimes \gamma_{\nu} \gamma_{\epsilon} \Psi^{\rho})
\nonumber\\
+ i~u^{\nu}~(- \bar{\Psi}_{\rho} t_{2}^{\epsilon} \otimes \gamma_{\nu} \gamma_{\epsilon} d^{\rho}\Psi^{\mu}
+ d^{\rho}\bar{\Psi}^{\mu} t_{2}^{\epsilon} \otimes \gamma_{\nu} \gamma_{\epsilon} \Psi_{\rho})
+ {i \over 2}~h~(\bar{\chi} t_{2}^{\epsilon} \otimes \gamma_{\nu} \gamma_{\epsilon} d^{\mu}\Psi^{\nu} 
- d^{\mu}\bar{\Psi}^{\nu} t_{2}^{\epsilon} \otimes \gamma_{\nu} \gamma_{\epsilon} \chi)
\nonumber\\
+ i~h_{\rho\sigma}~(- \bar{\chi} t_{2}^{\epsilon} \otimes \gamma^{\rho} \gamma_{\epsilon} d^{\mu}\Psi^{\sigma}
+ d^{\mu}\bar{\Psi}^{\rho} t_{2}^{\epsilon} \otimes \gamma^{\sigma} \gamma_{\epsilon} \chi)
+ i~h^{\mu\nu}~(d_{\rho}\bar{\chi} t_{2}^{\epsilon} \otimes \gamma_{\nu} \gamma_{\epsilon} \Psi^{\rho}
- \bar{\Psi}^{\rho} t_{2}^{\epsilon} \otimes \gamma_{\nu} \gamma_{\epsilon} d_{\rho}\chi)
\nonumber\\
+ {i \over 2}~h~(d^{\mu}~\bar{\chi} t_{2}^{\epsilon} \otimes \gamma_{\nu} \gamma_{\epsilon} \Psi^{\nu}
- \bar{\Psi}^{\nu} t_{2}^{\epsilon} \otimes \gamma_{\nu} \gamma_{\epsilon} d^{\mu}\chi)
- i~h_{\rho\sigma}~(d^{\mu}\bar{\chi} t_{2}^{\epsilon} \otimes \gamma^{\rho} \gamma_{\epsilon} \Psi^{\sigma}
- \bar{\Psi}_{\rho} t_{2}^{\epsilon} \otimes \gamma_{\sigma} \gamma_{\epsilon} d^{\mu}\chi)
\nonumber\\
-  {i \over 2}~d^{\mu}h~(\bar{\chi} t_{2}^{\epsilon} \otimes \gamma_{\nu} \gamma_{\epsilon} \Psi^{\nu}
- \bar{\Psi}^{\nu} t_{2}^{\epsilon} \otimes \gamma_{\nu} \gamma_{\epsilon} \chi)
+ i~d^{\mu}h^{\rho\sigma}~(\bar{\chi} t_{2}^{\epsilon} \otimes \gamma_{\rho} \gamma_{\epsilon} \Psi_{\sigma}
- \bar{\Psi}_{\rho} t_{2}^{\epsilon} \otimes \gamma_{\sigma} \gamma_{\epsilon} \chi)
\nonumber\\
- i \tilde{u}^{\nu}~(\bar{\chi} t_{2}^{\epsilon} \otimes \gamma_{\nu} \gamma_{\epsilon} d^{\mu}\chi
+ d^{\mu}\bar{\chi} t_{2}^{\epsilon} \otimes \gamma_{\nu} \gamma_{\epsilon} \chi)
+ i~d^{\mu}\tilde{u}^{\nu}~\bar{\chi} t_{2}^{\epsilon} \otimes \gamma_{\nu} \gamma_{\epsilon} \chi
\nonumber\\
+ {i\over 2}~u^{\nu}~(\bar{\tilde{\chi}} t_{2}^{\epsilon} \otimes \gamma_{\nu} \gamma_{\epsilon} d^{\mu}\chi
- d^{\mu}\bar{\chi} t_{2}^{\epsilon} \otimes \gamma_{\nu} \gamma_{\epsilon} \tilde{\chi})
+ {i\over 2}~u^{\nu}~(d^{\mu}\bar{\tilde{\chi}} t_{2}^{\epsilon} \otimes \gamma_{\nu} \gamma_{\epsilon} \chi
- \bar{\chi} t_{2}^{\epsilon} \otimes \gamma_{\nu} \gamma_{\epsilon} d^{\mu}\tilde{\chi})
\nonumber\\
- {i\over 2}~d^{\mu}u^{\nu}~(\bar{\tilde{\chi}} t_{2}^{\epsilon} \otimes \gamma_{\nu} \gamma_{\epsilon} \chi
- \bar{\chi} t_{2}^{\epsilon} \otimes \gamma_{\nu} \gamma_{\epsilon} \tilde{\chi}), \qquad
\eea
\bea
t_{2}^{\mu\nu} = {i \over 2}~(u^{\nu}~d^{\rho}\bar{\chi} t_{2}^{\epsilon} \otimes \gamma^{\mu} \gamma_{\epsilon} \Psi_{\rho}
- u^{\nu}~\Psi^{\rho} t_{2}^{\epsilon} \otimes \gamma^{\mu} \gamma_{\epsilon} d^{\rho}\chi
+ u^{\rho}~d^{\mu}\bar{\chi} t_{2}^{\epsilon} \otimes \gamma^{\nu} \gamma_{\epsilon} \Psi_{\rho}
\nonumber\\
+ u^{\rho}~\bar{\chi} t_{2}^{\epsilon} \otimes \gamma^{\nu} \gamma_{\epsilon} d^{\mu}\Psi_{\rho}
- d^{\mu}u^{\rho}~\bar{\chi} t_{2}^{\epsilon} \otimes \gamma^{\nu} \gamma_{\epsilon} \Psi_{\rho}
- u^{\rho}~d^{\nu}\bar{\chi} t_{2}^{\epsilon} \otimes \gamma^{\mu} \gamma_{\epsilon} \Psi_{\rho}
\nonumber\\
- u^{\rho}~\bar{\chi} t_{2}^{\epsilon} \otimes \gamma^{\mu} \gamma_{\epsilon} d^{\nu}\Psi_{\rho}
+ d^{\nu}u^{\rho}~\bar{\chi} t_{2}^{\epsilon} \otimes \gamma^{\mu} \gamma_{\epsilon} \Psi_{\rho}
- u^{\rho}~d^{\nu}\bar{\chi} t_{2}^{\epsilon} \otimes \gamma_{\rho} \gamma_{\epsilon} \Psi^{\mu}
\nonumber\\
- u^{\rho}~\bar{\chi} t_{2}^{\epsilon} \otimes \gamma_{\rho} \gamma_{\epsilon} d^{\nu}\Psi^{\mu}
+ d^{\nu}u^{\rho}~\bar{\chi} t_{2}^{\epsilon} \otimes \gamma_{\rho} \gamma_{\epsilon} \Psi^{\mu}
- u^{\rho}~d^{\mu}\Psi^{\nu} t_{2}^{\epsilon} \otimes \gamma^{\mu} \gamma_{\epsilon} \chi
\nonumber\\
- u^{\rho}~\Psi^{\nu} t_{2}^{\epsilon} \otimes \gamma_{\rho} \gamma_{\epsilon} d^{\mu}\chi
- d^{\nu}u^{\rho}~\Psi^{\mu} t_{2}^{\epsilon} \otimes \gamma_{\rho} \gamma_{\epsilon} \chi)
\nonumber\\
+ i~( h^{\mu\rho}~\bar{\chi} t_{2}^{\epsilon} \otimes \gamma_{\rho} \gamma_{\epsilon} d^{\nu}\chi
+ h^{\mu\rho}~d^{\nu}\bar{\chi} t_{2}^{\epsilon} \otimes \gamma_{\rho} \gamma_{\epsilon} \chi
+ d^{\mu}h^{\nu\rho}~\bar{\chi} t_{2}^{\epsilon} \otimes \gamma_{\rho} \gamma_{\epsilon} \chi) 
\nonumber\\
- (\mu \leftrightarrow \nu),
\eea
and
\bea
t_{2}^{\mu\nu\rho} = {i \over 2}~(- u^{\mu}~d^{\nu}\bar{\chi} t_{2}^{\epsilon} \otimes \gamma^{\rho} \gamma_{\epsilon} \chi
- u^{\nu}~d^{\rho}\chi t_{2}^{\epsilon} \otimes \gamma^{\mu} \gamma_{\epsilon} d^{\rho}\chi
- u^{\rho}~d^{\mu}\bar{\chi} t_{2}^{\epsilon} \otimes \gamma^{\nu} \gamma_{\epsilon} \chi
\nonumber\\
+ u^{\nu}~d^{\mu}\bar{\chi} t_{2}^{\epsilon} \otimes \gamma^{\nu} \gamma_{\epsilon} \chi
+ u^{\rho}~d^{\nu}\bar{\chi} t_{2}^{\epsilon} \otimes \gamma^{\mu} \gamma_{\epsilon} \chi
+ u^{\mu}~d^{\rho}\bar{\chi} t_{2}^{\epsilon} \otimes \gamma^{\nu} \gamma_{\epsilon} \chi
\nonumber\\
+ u^{\mu}~\bar{\chi} t_{2}^{\epsilon} \otimes \gamma^{\mu} \gamma_{\epsilon} d^{\rho}\chi
+ u^{\nu}~\bar{\chi} t_{2}^{\epsilon} \otimes \gamma^{\nu} \gamma_{\epsilon} d^{\mu}\chi
+ u^{\rho}~\bar{\chi} t_{2}^{\epsilon} \otimes \gamma_{\rho} \gamma_{\epsilon} d^{\nu}\chi
\nonumber\\
- u^{\nu}~\bar{\chi} t_{2}^{\epsilon} \otimes \gamma^{\mu} \gamma_{\epsilon} d^{\rho}\chi
- u^{\rho}~\bar{\chi} t_{2}^{\epsilon} \otimes \gamma^{\nu} \gamma_{\epsilon} d^{\mu}\chi
- u^{\mu}~\chi t_{2}^{\epsilon} \otimes \gamma^{\rho} \gamma_{\epsilon} d^{\nu}\chi
\nonumber\\
+ d^{\rho}u^{\nu}~\chi t_{2}^{\epsilon} \otimes \gamma^{\mu} \gamma_{\epsilon} \chi
+ d^{\mu}u^{\rho}~\chi t_{2}^{\epsilon} \otimes \gamma^{\nu} \gamma_{\epsilon} \chi
+ d^{\nu}u^{\mu}~\chi t_{2}^{\epsilon} \otimes \gamma^{\rho} \gamma_{\epsilon} \chi
\nonumber\\
- d^{\rho}u^{\mu}~\chi t_{2}^{\epsilon} \otimes \gamma^{\nu} \gamma_{\epsilon} \chi
- d^{\mu}u^{\nu}~\chi t_{2}^{\epsilon} \otimes \gamma^{\rho} \gamma_{\epsilon} \chi
- d^{\nu}u^{\rho}~\chi t_{2}^{\epsilon} \otimes \gamma^{\mu} \gamma_{\epsilon} \chi).
\eea
\label{3/2+2}
\end{thm}

{\bf Proof:} It is similar to the previous case only the list of terms $T$ and $T^{\mu}$ is longer. We first consider the 
case
$
\omega = 4.
$
We have
\be
T_{\omega = 4} = \sum T_{i}
\ee
\bea
T_{1} = h_{\mu\nu}~\bar{\Psi}^{\mu} t_{1}^{\epsilon} \otimes \gamma_{\epsilon} \Psi^{\nu}, \qquad
T_{2} = h~\bar{\Psi}^{\mu} t_{2}^{\epsilon} \otimes \gamma_{\epsilon} \Psi_{\mu}, 
\nonumber\\
T_{3} = h_{\mu\nu}~\bar{\Psi}^{\mu} t_{3}^{\epsilon} \otimes \gamma^{\nu}\gamma_{\rho} \gamma_{\epsilon} \Psi^{\rho}, \qquad
T_{4} = h_{\mu\nu}~\bar{\Psi}^{\rho} t_{4}^{\epsilon} \otimes \gamma_{\rho}\gamma^{\mu} \gamma_{\epsilon} \Psi^{\nu}, 
\nonumber\\
T_{5} = h~\bar{\Psi}^{\mu} t_{5}^{\epsilon} \otimes \gamma_{\mu}\gamma_{\nu} \gamma_{\epsilon} \Psi^{\nu}, \qquad
T_{6} = h~\bar{\tilde{\chi}} t_{6}^{\epsilon} \otimes \gamma_{\epsilon} \chi, 
\nonumber\\
T_{7} = h~\bar{\chi} t_{7}^{\epsilon} \otimes \gamma_{\epsilon} \tilde{\chi}, \qquad
T_{8} = u_{\mu}~\bar{\Psi}^{\mu} t_{8}^{\epsilon} \otimes \gamma_{\epsilon} \tilde{\chi}, 
\nonumber\\
T_{9} = u_{\mu}~\bar{\tilde{\chi}} t_{9}^{\epsilon} \otimes \gamma_{\epsilon} \Psi^{\mu}, \qquad
T_{10} = u_{\mu}~\bar{\Psi}_{\nu} t_{10}^{\epsilon} \otimes \gamma^{\nu}\gamma^{\mu} \gamma_{\epsilon} \tilde{\chi}, 
\nonumber\\
T_{11} = u_{\mu}~\bar{\tilde{\chi}} t_{11}^{\epsilon} \otimes \gamma^{\mu}\gamma^{\nu} \gamma_{\epsilon} \Psi_{\nu}, \qquad
T_{12} = \tilde{u}_{\mu}~\bar{\Psi}^{\mu} t_{12}^{\epsilon} \otimes \gamma_{\epsilon} \chi, 
\nonumber\\
T_{13} = \tilde{u}_{\mu}~\bar{\chi} t_{13}^{\epsilon} \otimes \gamma_{\epsilon} \Psi^{\mu}, \qquad
T_{14} = \tilde{u}_{\mu}~\bar{\Psi}_{\nu} t_{14}^{\epsilon} \otimes \gamma^{\nu}\gamma^{\mu} \gamma_{\epsilon} \chi, 
\nonumber\\
T_{15} = \tilde{u}_{\mu}~\bar{\chi} t_{15}^{\epsilon} \otimes \gamma^{\mu}\gamma^{\nu} \gamma_{\epsilon} \Psi^{\nu}.
\eea
Also
\be
T^{\mu}_{\omega = 4} = \sum T^{\mu}_{i}
\ee
where
\bea
T_{1}^{\mu} = u^{\mu}~\bar{\Psi}^{\nu} s_{1}^{\epsilon} \otimes \gamma_{\epsilon} \Psi_{\nu}, \qquad
T_{2}^{\mu} = u^{\mu}~\bar{\Psi}^{\nu} s_{2}^{\epsilon} \otimes \gamma_{\nu}\gamma_{\rho} \gamma_{\epsilon} \Psi^{\rho}, 
\nonumber\\
T_{3}^{\mu} = u^{\nu}~\bar{\Psi}^{\mu} s_{3}^{\epsilon} \otimes \gamma_{\epsilon} \Psi_{\nu}, \qquad
T_{4}^{\mu} = u^{\nu}~\bar{\Psi}_{\nu} s_{4}^{\epsilon} \otimes \gamma_{\epsilon} \Psi^{\mu}, 
\nonumber\\
T_{5}^{\mu} = u^{\nu}~\bar{\Psi}^{\mu} s_{5}^{\epsilon} \otimes \gamma_{\nu}\gamma_{\rho} \gamma_{\epsilon} \Psi^{\rho}, \qquad
T_{6}^{\mu} = u^{\nu}~\bar{\Psi}^{\rho} s_{6}^{\epsilon} \otimes \gamma_{\rho}\gamma_{\nu} \gamma_{\epsilon} \Psi^{\mu}, 
\nonumber\\
T_{7}^{\mu} = u^{\nu}~\bar{\Psi}_{\nu} s_{7}^{\epsilon} \otimes \gamma^{\mu}\gamma_{\rho} \gamma_{\epsilon} \Psi^{\rho}, \qquad
T_{8}^{\mu} = u^{\nu}~\bar{\Psi}_{\rho} s_{8}^{\epsilon} \otimes \gamma^{\rho}\gamma^{\mu} \gamma_{\epsilon} \Psi_{\nu}, 
\nonumber\\
T_{9}^{\mu} = h^{\mu\nu}~\bar{\chi} s_{9}^{\epsilon} \otimes \gamma_{\nu}\gamma_{\rho} \gamma_{\epsilon} \Psi^{\rho}, \qquad
T_{10}^{\mu} = h^{\mu\nu}~\bar{\chi} s_{10}^{\epsilon} \otimes \gamma_{\epsilon} \Psi_{\nu}, 
\nonumber\\
T_{11}^{\mu} = h~\bar{\chi} s_{11}^{\epsilon} \otimes \gamma^{\mu}\gamma_{\nu} \gamma_{\epsilon} \Psi^{\nu}, \qquad
T_{12}^{\mu} = h~\bar{\chi} s_{12}^{\epsilon} \otimes \gamma_{\epsilon} \Psi^{\mu}, 
\nonumber\\
T_{13}^{\mu} = h_{\mu\nu}~\bar{\Psi}^{\rho} s_{13}^{\epsilon} \otimes \gamma_{\rho}\gamma_{\nu} \gamma_{\epsilon} \chi, \qquad
T_{14}^{\mu} = h_{\mu\nu}~\bar{\Psi}_{\nu} s_{14}^{\epsilon} \otimes \gamma_{\epsilon} \chi, 
\nonumber\\
T_{15}^{\mu} = h~\bar{\Psi}^{\nu} s_{15}^{\epsilon} \otimes \gamma_{\nu}\gamma^{\mu} \gamma_{\epsilon} \chi, \qquad
T_{16}^{\mu} = h~\bar{\Psi}^{\mu} s_{16}^{\epsilon} \otimes \gamma_{\epsilon} \chi, 
\nonumber\\
T_{17}^{\mu} = \tilde{u}^{\mu}~\bar{\chi} s_{17}^{\epsilon} \otimes \gamma_{\epsilon} \chi, \qquad
T_{18}^{\mu} = \tilde{u}_{\nu}~\bar{\chi} s_{18}^{\epsilon} \otimes \gamma^{\mu}\gamma^{\nu} \gamma_{\epsilon} \chi, 
\nonumber \\
T_{19}^{\mu} = u^{\mu}~\bar{\tilde{\chi}} s_{19}^{\epsilon} \otimes \gamma_{\epsilon} \chi, \qquad
T_{20}^{\mu} = u^{\mu}~\bar{\chi} s_{20}^{\epsilon} \otimes \gamma_{\epsilon} \tilde{\chi}, 
\nonumber\\
T_{21}^{\mu} = u_{\nu}~\bar{\tilde{\chi}} s_{21}^{\epsilon} \otimes \gamma^{\mu}\gamma^{\nu} \gamma_{\epsilon} \chi, \qquad
T_{22}^{\mu} = u_{\nu}~\bar{\chi} s_{22}^{\epsilon} \otimes \gamma^{\mu}\gamma^{\nu} \gamma_{\epsilon} \tilde{\chi}.
\eea
We substitute these expressions in (\ref{gauge1a}) and, as in the case of pure gravity, we do not find solutions for
$
\omega = 4.
$

We investigate now the case
$
\omega = 5.
$
We have 
\be
T_{\omega = 5} = \sum T_{i}
\ee
where
\bea
T_{1} = h_{\mu\nu}~\bar{\Psi}^{\mu} t_{1}^{\epsilon} \otimes \gamma^{\nu} \gamma_{\epsilon} d_{\rho}\Psi^{\rho}, \qquad
T_{2} = h_{\mu\nu}~\bar{\Psi}^{\mu} t_{2}^{\epsilon} \otimes \gamma_{\rho} \gamma_{\epsilon} d^{\nu}\Psi^{\rho}, 
\nonumber\\
T_{3} = h_{\mu\nu}~\bar{\Psi}_{\rho} t_{3}^{\epsilon} \otimes \gamma^{\mu} \gamma_{\epsilon} d^{\nu}\Psi^{\rho}, \qquad
T_{4} = h_{\mu\nu}~\bar{\Psi}_{\rho} t_{4}^{\epsilon} \otimes \gamma^{\mu} \gamma_{\epsilon} d^{\rho}\Psi^{\nu}, 
\nonumber\\
T_{5} = h_{\mu\nu}~\bar{\Psi}_{\rho} t_{5}^{\epsilon} \otimes \gamma^{\rho} \gamma_{\epsilon} d^{\mu}\Psi^{\nu}, \qquad
T_{6} = h~\bar{\Psi}^{\mu} t_{6}^{\epsilon} \otimes \gamma_{\mu} \gamma_{\epsilon} d^{\nu}\Psi_{\nu}, 
\nonumber\\
T_{7} = h~\bar{\Psi}^{\mu} t_{7}^{\epsilon} \otimes \gamma^{\nu} \gamma_{\epsilon} d_{\mu}\Psi_{\nu}, \qquad
T_{8} = h_{\mu\nu}~d^{\rho}\bar{\Psi}_{\rho} t_{8}^{\epsilon} \otimes \gamma^{\mu} \gamma_{\epsilon} \Psi^{\nu}, 
\nonumber\\
T_{9} = h_{\mu\nu}~d^{\mu}\bar{\Psi}^{\rho} t_{9}^{\epsilon} \otimes \gamma_{\rho} \gamma_{\epsilon} \Psi^{\nu}, \qquad
T_{10} = h_{\mu\nu}~d^{\mu}\bar{\Psi}^{\rho} t_{10}^{\epsilon} \otimes \gamma^{\nu} \gamma_{\epsilon} \Psi_{\rho}, 
\nonumber\\
T_{11} = h_{\mu\nu}~d^{\rho}\bar{\Psi}^{\mu} t_{11}^{\epsilon} \otimes \gamma^{\nu} \gamma_{\epsilon} \Psi_{\rho}, \qquad
T_{12} = h_{\mu\nu}~d^{\mu}\bar{\Psi}^{\nu} t_{12}^{\epsilon} \otimes \gamma^{\mu} \gamma_{\epsilon} \Psi_{\rho}, 
\nonumber\\
T_{13} = h~d_{\mu}\bar{\Psi}^{\mu} t_{13}^{\epsilon} \otimes \gamma^{\nu} \gamma_{\epsilon} \Psi_{\nu}, \qquad
T_{14} = h~d^{\mu}\bar{\Psi}^{\nu} t_{14}^{\epsilon} \otimes \gamma_{\nu} \gamma_{\epsilon} \Psi_{\mu}, 
\nonumber\\
T_{15} = h_{\mu\nu}~\bar{\Psi}^{\rho} t_{15}^{\epsilon} \otimes 
\gamma_{\rho}\gamma^{\mu}\gamma_{\sigma} \gamma_{\epsilon} d^{\nu}\Psi^{\sigma}, \qquad
T_{16} = h_{\mu\nu}~d^{\mu}\bar{\Psi}^{\rho} t_{16}^{\epsilon} \otimes 
\gamma_{\rho}\gamma^{\nu}\gamma_{\sigma} \gamma_{\epsilon} \Psi^{\sigma}, 
\nonumber\\
T_{17} = h_{\mu\nu}~\bar{\tilde{\chi}} t_{17}^{\epsilon} \otimes \gamma^{\mu} \gamma_{\epsilon} d^{\nu}\chi, \qquad
T_{18} = h_{\mu\nu}~d^{\mu}\bar{\tilde{\chi}} t_{18}^{\epsilon} \otimes \gamma^{\nu} \gamma_{\epsilon} \chi, 
\nonumber \\
T_{19} = h_{\mu\nu}~\bar{\chi} t_{19}^{\epsilon} \otimes \gamma^{\mu} \gamma_{\epsilon} d^{\nu}\tilde{\chi}, \qquad
T_{20} = h_{\mu\nu}~d^{\mu}\bar{\chi} t_{20}^{\epsilon} \otimes \gamma^{\nu} \gamma_{\epsilon} \tilde{\chi}, 
\nonumber\\
T_{21} = u_{\mu}~d^{\mu}\bar{\Psi}^{\nu} t_{21}^{\epsilon} \otimes \gamma_{\nu} \gamma_{\epsilon} \tilde{\chi}, \qquad
T_{22} = u_{\mu}~d_{\nu}\bar{\Psi}^{\nu} t_{22}^{\epsilon} \otimes \gamma^{\mu} \gamma_{\epsilon} \tilde{\chi}, 
\nonumber\\
T_{23} = u_{\mu}~\bar{\Psi}_{\nu} t_{23}^{\epsilon} \otimes \gamma^{\mu} \gamma_{\epsilon} d^{\nu}\tilde{\chi}, \qquad
T_{24} = u_{\mu}~\bar{\Psi}_{\nu} t_{24}^{\epsilon} \otimes \gamma^{\nu} \gamma_{\epsilon} d^{\mu}\tilde{\chi},
\nonumber\\
T_{25} = u_{\mu}~d^{\mu}\bar{\tilde{\chi}} t_{25}^{\epsilon} \otimes \gamma_{\rho} \gamma_{\epsilon} \Psi^{\rho}, \qquad
T_{26} = u_{\mu}~d_{\nu}\bar{\tilde{\chi}} t_{26}^{\epsilon} \otimes \gamma^{\mu} \gamma_{\epsilon} \Psi^{\nu},
\nonumber\\
T_{27} = u_{\mu}~\bar{\tilde{\chi}} t_{27}^{\epsilon} \otimes \gamma^{\mu} \gamma_{\epsilon} d_{\nu}\Psi^{\nu}, \qquad
T_{28} = u_{\mu}~\bar{\tilde{\chi}} t_{28}^{\epsilon} \otimes \gamma_{\nu} \gamma_{\epsilon} d^{\mu}\Psi^{\nu},
\nonumber\\
T_{29} = \tilde{u}_{\mu}~\bar{\Psi}_{\nu} t_{29}^{\epsilon} \otimes \gamma^{\mu} \gamma_{\epsilon} d^{\nu}\chi, \qquad
T_{30} = \tilde{u}_{\mu}~\bar{\Psi}_{\nu} t_{30}^{\epsilon} \otimes \gamma^{\nu} \gamma_{\epsilon} d^{\mu}\chi,
\nonumber\\
T_{31} = \tilde{u}_{\mu}~d_{\nu}\bar{\Psi}^{\nu} t_{31}^{\epsilon} \otimes \gamma^{\mu} \gamma_{\epsilon} \chi, \qquad
T_{32} = \tilde{u}_{\mu}~d^{\mu}\bar{\Psi}^{\nu} t_{32}^{\epsilon} \otimes \gamma_{\nu} \gamma_{\epsilon} \chi,
\nonumber\\
T_{33} = \tilde{u}_{\mu}~\bar{\chi} t_{33}^{\epsilon} \otimes \gamma^{\mu} \gamma_{\epsilon} d_{\nu}\Psi^{\nu}, \qquad
T_{34} = \tilde{u}_{\mu}~\bar{\chi} t_{34}^{\epsilon} \otimes \gamma_{\nu} \gamma_{\epsilon} d^{\mu}\Psi^{\nu},
\nonumber\\
T_{35} = \tilde{u}_{\mu}~d_{\nu}\bar{\chi} t_{35}^{\epsilon} \otimes \gamma^{\mu} \gamma_{\epsilon} \Psi^{\nu}, \qquad
T_{36} = \tilde{u}_{\mu}~d^{\mu}\bar{\chi} t_{36}^{\epsilon} \otimes \gamma_{\nu} \gamma_{\epsilon} \Psi^{\nu}
\eea
and we have eliminated expressions of the type
$
d_{\rho}h_{\mu\nu} \bar{\Psi} \Psi, d_{\mu}u_{\nu} \bar{\Psi} \tilde{\chi}, d_{\mu}u_{\nu} \bar{\tilde{\chi}} \Psi,
d_{\mu}\tilde{u}_{\nu} \bar{\Psi} \chi, d_{\mu}\tilde{u}_{\nu} \bar{\chi} \Psi
$
using total derivatives. We also have
\be
T^{\mu}_{\omega = 5} = \sum T^{\mu}_{i}
\ee
where
\bea
T_{1}^{\mu} = u^{\mu}~\bar{\Psi}^{\nu} s_{1}^{\epsilon} \otimes \gamma_{\nu} \gamma_{\epsilon} d_{\rho}\Psi^{\rho}, \qquad
T_{2}^{\mu} = u^{\mu}~\bar{\Psi}^{\nu} s_{2}^{\epsilon} \otimes \gamma_{\rho} \gamma_{\epsilon} d_{\nu}\Psi^{\rho}, 
\nonumber\\
T_{3}^{\mu} = u^{\nu}~\bar{\Psi}^{\mu} s_{3}^{\epsilon} \otimes \gamma_{\nu} \gamma_{\epsilon} d_{\rho}\Psi^{\rho}, \qquad
T_{4}^{\mu} = u^{\nu}~\bar{\Psi}^{\mu} s_{4}^{\epsilon} \otimes \gamma_{\rho} \gamma_{\epsilon} d_{\nu}\Psi^{\rho}, 
\nonumber\\
T_{5}^{\mu} = u^{\nu}~\bar{\Psi}_{\nu} s_{5}^{\epsilon} \otimes \gamma^{\mu} \gamma_{\epsilon} d_{\rho}\Psi^{\rho}, \qquad
T_{6}^{\mu} = u^{\nu}~\bar{\Psi}^{\rho} s_{6}^{\epsilon} \otimes \gamma^{\mu} \gamma_{\epsilon} d_{\nu}\Psi_{\rho}, 
\nonumber\\
T_{7}^{\mu} = u^{\nu}~\bar{\Psi}^{\rho} s_{7}^{\epsilon} \otimes \gamma^{\mu} \gamma_{\epsilon} d_{\rho}\Psi_{\nu}, \qquad
T_{8}^{\mu} = u^{\nu}~\bar{\Psi}_{\nu} s_{8}^{\epsilon} \otimes \gamma_{\rho} \gamma_{\epsilon} d^{\mu}\Psi^{\rho}, 
\nonumber\\
T_{9}^{\mu} = u^{\nu}~\bar{\Psi}_{\rho} s_{9}^{\epsilon} \otimes \gamma_{\nu} \gamma_{\epsilon} d^{\mu}\Psi^{\rho}, \qquad
T_{10}^{\mu} = u^{\nu}~\bar{\Psi}_{\rho} s_{10}^{\epsilon} \otimes \gamma^{\rho} \gamma_{\epsilon} d^{\mu}\Psi_{\nu}, 
\nonumber\\
T_{11}^{\mu} = u^{\nu}~\bar{\Psi}_{\rho} s_{11}^{\epsilon} \otimes \gamma_{\nu} \gamma_{\epsilon} d^{\rho}\Psi^{\mu}, \qquad
T_{12}^{\mu} = u^{\nu}~\bar{\Psi}^{\rho} s_{12}^{\epsilon} \otimes \gamma_{\rho} \gamma_{\epsilon} d_{\nu}\Psi^{\mu}, 
\nonumber\\
T_{13}^{\mu} = u_{\nu}~\bar{\Psi}^{\rho} s_{13}^{\epsilon} \otimes 
\gamma_{\rho}\gamma^{\nu}\gamma_{\sigma} \gamma_{\epsilon} d^{\mu}\Psi^{\sigma}, \qquad
T_{14}^{\mu} = u_{\nu}~\bar{\Psi}^{\rho} s_{14}^{\epsilon} \otimes 
\gamma_{\rho}\gamma^{\mu}\gamma_{\sigma} \gamma_{\epsilon} d^{\nu}\Psi^{\sigma}, 
\nonumber\\
T_{15}^{\mu} = u^{\mu}~d_{\nu}\bar{\Psi}^{\nu} s_{15}^{\epsilon} \otimes \gamma_{\rho} \gamma_{\epsilon} \Psi^{\rho}, \qquad
T_{16}^{\mu} = u^{\mu}~d^{\nu}\bar{\Psi}^{\rho} s_{16}^{\epsilon} \otimes \gamma_{\rho} \gamma_{\epsilon} \Psi_{\nu}, 
\nonumber\\
T_{17}^{\mu} = u^{\nu}~d^{\mu}\bar{\Psi}_{\nu} s_{17}^{\epsilon} \otimes \gamma_{\rho} \gamma_{\epsilon} \Psi^{\rho}, \qquad
T_{18}^{\mu} = u^{\nu}~d^{\mu}\bar{\Psi}_{\rho} s_{18}^{\epsilon} \otimes \gamma_{\nu} \gamma_{\epsilon} \Psi^{\rho}, 
\nonumber \\
T_{19}^{\mu} = u^{\nu}~d^{\mu}\bar{\Psi}_{\rho} s_{19}^{\epsilon} \otimes \gamma^{\rho} \gamma_{\epsilon} \Psi_{\nu}, \qquad
T_{20}^{\mu} = u^{\nu}~d_{\nu}\bar{\Psi}^{\mu} s_{20}^{\epsilon} \otimes \gamma_{\rho} \gamma_{\epsilon} \Psi^{\rho}, 
\nonumber\\
T_{21}^{\mu} = u^{\nu}~d^{\rho}\bar{\Psi}^{\mu} s_{21}^{\epsilon} \otimes \gamma_{\nu} \gamma_{\epsilon} \Psi_{\rho}, \qquad
T_{22}^{\mu} = u^{\nu}~d_{\nu}\bar{\Psi}_{\rho} s_{22}^{\epsilon} \otimes \gamma^{\mu} \gamma_{\epsilon} \Psi^{\rho}, 
\nonumber\\
T_{23}^{\mu} = u^{\nu}~d_{\rho}\bar{\Psi}_{\nu} s_{23}^{\epsilon} \otimes \gamma^{\mu} \gamma_{\epsilon} \Psi^{\rho}, \qquad
T_{24}^{\mu} = u^{\nu}~d_{\rho}\bar{\Psi}^{\rho} s_{24}^{\epsilon} \otimes \gamma^{\mu} \gamma_{\epsilon} \Psi_{\nu},
\nonumber\\
T_{25}^{\mu} = u^{\nu}~d_{\nu}\bar{\Psi}_{\rho} s_{25}^{\epsilon} \otimes \gamma^{\rho} \gamma_{\epsilon} \Psi^{\mu}, \qquad
T_{26}^{\mu} = u^{\nu}~d_{\rho}\bar{\Psi}^{\rho} s_{26}^{\epsilon} \otimes \gamma_{\nu} \gamma_{\epsilon} \Psi^{\mu},
\nonumber\\
T_{27}^{\mu} = u_{\nu}~d^{\mu}\bar{\Psi}^{\rho} s_{27}^{\epsilon} \otimes 
\gamma_{\rho}\gamma^{\nu}\gamma_{\sigma} \gamma_{\epsilon} \Psi^{\sigma}, \qquad
T_{28}^{\mu} = u_{\nu}~d^{\nu}\bar{\Psi}^{\rho} s_{28}^{\epsilon} \otimes 
\gamma_{\rho}\gamma^{\mu}\gamma_{\sigma} \gamma_{\epsilon} \Psi^{\sigma},
\nonumber\\
T_{29}^{\mu} = d^{\mu}u^{\nu}~\bar{\Psi}_{\nu} s_{29}^{\epsilon} \otimes \gamma_{\rho} \gamma_{\epsilon} \Psi^{\rho}, \qquad
T_{30}^{\mu} = d^{\mu}u^{\nu}~\bar{\Psi}_{\rho} s_{30}^{\epsilon} \otimes \gamma_{\nu} \gamma_{\epsilon} \Psi^{\rho},
\nonumber\\
T_{31}^{\mu} = d^{\mu}u^{\nu}~\bar{\Psi}_{\rho} s_{31}^{\epsilon} \otimes \gamma^{\rho} \gamma_{\epsilon} \Psi_{\nu}, \qquad
T_{32}^{\mu} = d^{\nu}u^{\mu}~\bar{\Psi}_{\nu} s_{32}^{\epsilon} \otimes \gamma_{\rho} \gamma_{\epsilon} \Psi^{\rho},
\nonumber\\
T_{33}^{\mu} = d^{\nu}u^{\mu}~\bar{\Psi}_{\rho} s_{33}^{\epsilon} \otimes \gamma_{\nu} \gamma_{\epsilon} \Psi^{\rho}, \qquad
T_{34}^{\mu} = d^{\nu}u^{\mu}~\bar{\Psi}_{\rho} s_{34}^{\epsilon} \otimes \gamma^{\rho} \gamma_{\epsilon} \Psi_{\nu},
\nonumber\\
T_{35}^{\mu} = d_{\nu}u^{\nu}~\bar{\Psi}^{\mu} s_{35}^{\epsilon} \otimes \gamma_{\rho} \gamma_{\epsilon} \Psi^{\rho}, \qquad
T_{36}^{\mu} = d_{\nu}u_{\rho}~\bar{\Psi}^{\mu} s_{36}^{\epsilon} \otimes \gamma^{\nu} \gamma_{\epsilon} \Psi^{\rho},
\nonumber\\
T_{37}^{\mu} = d_{\nu}u_{\rho}~\bar{\Psi}^{\mu} s_{37}^{\epsilon} \otimes \gamma^{\rho} \gamma_{\epsilon} \Psi^{\nu}, \qquad
T_{38}^{\mu} = d_{\nu}u^{\nu}~\bar{\Psi}^{\rho} s_{38}^{\epsilon} \otimes \gamma^{\mu} \gamma_{\epsilon} \Psi_{\rho},
\nonumber\\
T_{39}^{\mu} = d_{\nu}u_{\rho}~\bar{\Psi}^{\nu} s_{39}^{\epsilon} \otimes \gamma^{\mu} \gamma_{\epsilon} \Psi^{\rho}, \qquad
T_{40}^{\mu} = d_{\nu}u_{\rho}~\bar{\Psi}^{\rho} s_{40}^{\epsilon} \otimes \gamma^{\mu} \gamma_{\epsilon} \Psi_{\nu},
\nonumber\\
T_{41}^{\mu} = d_{\nu}u^{\nu}~\bar{\Psi}^{\rho} s_{41}^{\epsilon} \otimes \gamma_{\rho} \gamma_{\epsilon} \Psi^{\mu}, \qquad
T_{42}^{\mu} = d_{\nu}u_{\rho}~\bar{\Psi}^{\nu} s_{42}^{\epsilon} \otimes \gamma^{\rho} \gamma_{\epsilon} \Psi_{\mu},
\nonumber\\
T_{43}^{\mu} = d_{\nu} u_{\rho}~\bar{\Psi}^{\rho} s_{43}^{\epsilon} \otimes \gamma^{\nu} \gamma_{\epsilon} \Psi^{\mu}, \qquad
T_{44}^{\mu} = d^{\mu}u^{\nu}~\bar{\Psi}^{\rho} s_{44}^{\epsilon} \otimes 
\gamma_{\rho}\gamma_{\nu}\gamma_{\sigma} \gamma_{\epsilon} \Psi^{\sigma},
\nonumber\\
T_{45}^{\mu} = d^{\nu}u^{\mu}~\bar{\Psi}^{\rho} s_{45}^{\epsilon} \otimes 
\gamma_{\rho}\gamma_{\nu}\gamma_{\sigma} \gamma_{\epsilon} \Psi^{\sigma},\qquad
T_{46}^{\mu} = d^{\nu}u^{\rho}~\bar{\Psi}^{\mu} s_{46}^{\epsilon} \otimes 
\gamma_{\nu}\gamma_{\rho}\gamma_{\sigma} \gamma_{\epsilon} \Psi^{\sigma},
\nonumber\\
T_{47}^{\mu} = d^{\nu}u^{\rho}~\bar{\Psi}_{\nu} s_{47}^{\epsilon} \otimes 
\gamma_{\rho}\gamma^{\mu}\gamma_{\sigma} \gamma_{\epsilon} \Psi^{\sigma},\qquad
T_{48}^{\mu} = d^{\nu}u^{\rho}~\bar{\Psi}_{\rho} s_{48}^{\epsilon} \otimes 
\gamma_{\nu}\gamma^{\mu}\gamma_{\sigma} \gamma_{\epsilon} \Psi^{\sigma},
\nonumber\\
T_{49}^{\mu} = d_{\sigma}u^{\sigma}~\bar{\Psi}^{\nu} s_{49}^{\epsilon} \otimes 
\gamma_{\nu}\gamma^{\mu}\gamma_{\rho} \gamma_{\epsilon} \Psi^{\rho},\qquad
T_{50}^{\mu} = d^{\nu}u^{\rho}~\bar{\Psi}^{\sigma} s_{50}^{\epsilon} \otimes 
\gamma_{\sigma}\gamma^{\mu}\gamma_{\rho} \gamma_{\epsilon} \Psi_{\nu},
\nonumber\\
T_{51}^{\mu} = d^{\nu}u^{\rho}~\bar{\Psi}^{\sigma} s_{51}^{\epsilon} \otimes 
\gamma_{\sigma}\gamma^{\mu}\gamma_{\nu} \gamma_{\epsilon} \Psi_{\rho},\qquad
T_{52}^{\mu} = d^{\nu}u^{\rho}~\bar{\Psi}^{\sigma} s_{52}^{\epsilon} \otimes 
\gamma_{\sigma}\gamma_{\nu}\gamma_{\rho} \gamma_{\epsilon} \Psi^{\mu},
\nonumber\\
T_{53}^{\mu} = h^{\mu\nu}~\bar{\chi} s_{53}^{\epsilon} \otimes \gamma_{\nu} \gamma_{\epsilon} d_{\rho}\Psi^{\rho},\qquad
T_{54}^{\mu} = h^{\mu\nu}~\bar{\chi} s_{54}^{\epsilon} \otimes \gamma_{\rho} \gamma_{\epsilon} d_{\nu}\Psi^{\rho},
\nonumber\\
T_{55}^{\mu} = h~\bar{\chi} s_{55}^{\epsilon} \otimes \gamma^{\mu} \gamma_{\epsilon} d_{\nu}\Psi^{\nu},\qquad
T_{56}^{\mu} = h_{\rho\sigma}~\bar{\chi} s_{56}^{\epsilon} \otimes \gamma^{\mu} \gamma_{\epsilon} d^{\rho}\Psi^{\sigma},
\nonumber\\
T_{57}^{\mu} = h~\bar{\chi} s_{57}^{\epsilon} \otimes \gamma_{\nu} \gamma_{\epsilon} d^{\mu}\Psi^{\nu},\qquad
T_{58}^{\mu} = h_{\rho\sigma}~\bar{\chi} s_{58}^{\epsilon} \otimes \gamma^{\rho} \gamma_{\epsilon} d^{\mu}\Psi^{\sigma},
\nonumber\\
T_{59}^{\mu} = h_{\rho\sigma}~\bar{\chi} s_{59}^{\epsilon} \otimes \gamma^{\rho} \gamma_{\epsilon} d^{\sigma}\Psi^{\mu},\qquad
T_{60}^{\mu} = h^{\mu\nu}~d_{\nu}\bar{\chi} s_{60}^{\epsilon} \otimes \gamma_{\rho} \gamma_{\epsilon} \Psi^{\rho},
\nonumber\\
T_{61}^{\mu} = h^{\mu\nu}~d_{\rho}\bar{\chi} s_{61}^{\epsilon} \otimes \gamma_{\nu} \gamma_{\epsilon} \Psi^{\rho},\qquad
T_{62}^{\mu} = h~d^{\mu}~\bar{\chi} s_{62}^{\epsilon} \otimes \gamma_{\nu} \gamma_{\epsilon} \Psi^{\nu},
\nonumber\\
T_{63}^{\mu} = h_{\rho\sigma}~d^{\mu}\bar{\chi} s_{63}^{\epsilon} \otimes \gamma^{\rho} \gamma_{\epsilon} \Psi^{\sigma},\qquad
T_{64}^{\mu} = h~d_{\nu}\bar{\chi} s_{64}^{\epsilon} \otimes \gamma^{\mu} \gamma_{\epsilon} \Psi^{\nu},
\nonumber\\
T_{65}^{\mu} = h_{\rho\sigma}~d^{\rho}\bar{\chi} s_{65}^{\epsilon} \otimes \gamma^{\mu} \gamma_{\epsilon} \Psi^{\sigma},\qquad
T_{66}^{\mu} = h_{\rho\sigma}~d^{\rho}\bar{\chi} s_{66}^{\epsilon} \otimes \gamma^{\sigma} \gamma_{\epsilon} \Psi^{\mu},
\nonumber\\
T_{67}^{\mu} = d^{\mu}h~\bar{\chi} s_{67}^{\epsilon} \otimes \gamma_{\nu} \gamma_{\epsilon} \Psi^{\nu},\qquad
T_{68}^{\mu} = d^{\mu}h^{\rho\sigma}~\bar{\chi} s_{68}^{\epsilon} \otimes \gamma_{\rho} \gamma_{\epsilon} \Psi_{\sigma},
\nonumber\\
T_{69}^{\mu} = d_{\nu}h^{\mu\nu}~\bar{\chi} s_{69}^{\epsilon} \otimes \gamma_{\rho} \gamma_{\epsilon} \Psi^{\rho},\qquad
T_{70}^{\mu} = d^{\rho}h^{\mu\nu}~\bar{\chi} s_{70}^{\epsilon} \otimes \gamma_{\nu} \gamma_{\epsilon} \Psi_{\rho},
\nonumber\\
T_{71}^{\mu} = d^{\rho}h^{\mu\nu}~\bar{\chi} s_{71}^{\epsilon} \otimes \gamma_{\rho} \gamma_{\epsilon} \Psi_{\nu},\qquad
T_{72}^{\mu} = d^{\nu}h~\bar{\chi} s_{72}^{\epsilon} \otimes \gamma_{\mu} \gamma_{\epsilon} \Psi^{\nu},
\nonumber\\
T_{73}^{\mu} = d_{\rho}h^{\nu\rho}~\bar{\chi} s_{73}^{\epsilon} \otimes \gamma^{\mu} \gamma_{\epsilon} \Psi_{\nu},\qquad
T_{74}^{\mu} = d^{\nu}h~\bar{\chi} s_{74}^{\epsilon} \otimes \gamma_{\nu} \gamma_{\epsilon} \Psi^{\mu},
\nonumber\\
T_{75}^{\mu} = d_{\rho}h^{\nu\rho}~\bar{\chi} s_{75}^{\epsilon} \otimes \gamma_{\nu} \gamma_{\epsilon} \Psi^{\mu},\qquad
T_{76}^{\mu} = h^{\mu\nu}~\bar{\Psi}^{\rho} s_{76}^{\epsilon} \otimes \gamma_{\nu} \gamma_{\epsilon} d_{\rho}\chi,
\nonumber\\
T_{77}^{\mu} = h^{\mu\nu}~\bar{\Psi}^{\rho} s_{77}^{\epsilon} \otimes \gamma_{\rho} \gamma_{\epsilon} d_{\nu}\chi, \qquad
T_{78}^{\mu} = h^{\rho\sigma}~\bar{\Psi}^{\mu} s_{78}^{\epsilon} \otimes \gamma_{\rho} \gamma_{\epsilon} d_{\sigma}\chi,
\nonumber\\
T_{79}^{\mu} = h~\bar{\Psi}^{\nu} s_{79}^{\epsilon} \otimes \gamma^{\mu} \gamma_{\epsilon} d_{\nu}\chi, \qquad
T_{80}^{\mu} = h^{\rho\sigma}~\bar{\Psi}_{\rho} s_{80}^{\epsilon} \otimes \gamma^{\mu} \gamma_{\epsilon} d_{\sigma}\chi,
\nonumber\\
T_{81}^{\mu} = h~\bar{\Psi}^{\nu} s_{81}^{\epsilon} \otimes \gamma_{\nu} \gamma_{\epsilon} d^{\mu}\chi, \qquad
T_{82}^{\mu} = h^{\rho\sigma}~\bar{\Psi}_{\rho} s_{82}^{\epsilon} \otimes \gamma_{\sigma} \gamma_{\epsilon} d^{\mu}\chi,
\nonumber\\
T_{83}^{\mu} = h^{\mu\nu}~d_{\nu}\bar{\Psi}_{\rho} s_{83}^{\epsilon} \otimes \gamma^{\rho} \gamma_{\epsilon} \chi, \qquad
T_{84}^{\mu} = h^{\mu\nu}~d_{\rho}\bar{\Psi}^{\rho} s_{84}^{\epsilon} \otimes \gamma_{\nu} \gamma_{\epsilon} \chi,
\nonumber\\
T_{85}^{\mu} = h~d^{\mu}\bar{\Psi}^{\nu} s_{85}^{\epsilon} \otimes \gamma_{\nu} \gamma_{\epsilon} \chi, \qquad
T_{86}^{\mu} = h_{\rho\sigma}~d^{\mu}\bar{\Psi}^{\rho} s_{86}^{\epsilon} \otimes \gamma^{\sigma} \gamma_{\epsilon} \chi,
\nonumber\\
T_{87}^{\mu} = h_{\rho\sigma}~d^{\rho}\bar{\Psi}^{\mu} s_{87}^{\epsilon} \otimes \gamma^{\sigma} \gamma_{\epsilon} \chi, \qquad
T_{88}^{\mu} = h~d_{\nu}\bar{\Psi}^{\nu} s_{88}^{\epsilon} \otimes \gamma^{\mu} \gamma_{\epsilon} \chi,
\nonumber\\
T_{89}^{\mu} = h_{\rho\sigma}~d^{\rho}\bar{\Psi}^{\sigma} s_{89}^{\epsilon} \otimes \gamma^{\mu} \gamma_{\epsilon} \chi, \qquad
T_{90}^{\mu} = d^{\mu}h~\bar{\Psi}^{\nu} s_{90}^{\epsilon} \otimes \gamma_{\nu} \gamma_{\epsilon} \chi,
\nonumber\\
T_{91}^{\mu} = d^{\mu}h^{\rho\sigma}~\bar{\Psi}_{\rho} s_{91}^{\epsilon} \otimes \gamma_{\sigma} \gamma_{\epsilon} \chi, \qquad
T_{92}^{\mu} = d_{\nu}h^{\mu\nu}~\bar{\Psi}_{\rho} s_{92}^{\epsilon} \otimes \gamma^{\rho} \gamma_{\epsilon} \chi,
\nonumber\\
T_{93}^{\mu} = d^{\rho}h^{\mu\nu}~\bar{\Psi}_{\rho} s_{93}^{\epsilon} \otimes \gamma_{\nu} \gamma_{\epsilon} \chi, \qquad
T_{94}^{\mu} = d^{\rho}h^{\mu\nu}~\bar{\Psi}_{\nu} s_{94}^{\epsilon} \otimes \gamma_{\rho} \gamma_{\epsilon} \chi,
\nonumber\\
T_{95}^{\mu} = d^{\nu}h~\bar{\Psi}^{\mu} s_{95}^{\epsilon} \otimes \gamma_{\nu} \gamma_{\epsilon} \chi, \qquad
T_{96}^{\mu} = d_{\rho}h^{\nu\rho}~\bar{\Psi}^{\mu} s_{96}^{\epsilon} \otimes \gamma_{\nu} \gamma_{\epsilon} \chi,
\nonumber\\
T_{97}^{\mu} = d_{\nu}h~\bar{\Psi}^{\nu} s_{97}^{\epsilon} \otimes \gamma^{\mu} \gamma_{\epsilon} \chi, \qquad
T_{98}^{\mu} = d_{\rho}h^{\nu\rho}~\bar{\Psi}_{\nu} s_{98}^{\epsilon} \otimes \gamma^{\mu} \gamma_{\epsilon} \chi,
\nonumber\\
T_{99}^{\mu} = \tilde{u}^{\nu}~\bar{\chi} s_{99}^{\epsilon} \otimes \gamma^{\mu} \gamma_{\epsilon} d_{\nu}\chi, \qquad
T_{100}^{\mu} = \tilde{u}^{\nu}~\bar{\chi} s_{100}^{\epsilon} \otimes \gamma_{\nu} \gamma_{\epsilon} d^{\mu}\chi,
\nonumber\\
T_{101}^{\mu} = \tilde{u}^{\nu}~d^{\mu}\bar{\chi} s_{101}^{\epsilon} \otimes \gamma_{\nu} \gamma_{\epsilon} \chi, \qquad
T_{102}^{\mu} = \tilde{u}^{\nu}~d_{\nu}\bar{\chi} s_{102}^{\epsilon} \otimes \gamma^{\mu} \gamma_{\epsilon} \chi,
\nonumber\\
T_{103}^{\mu} = d^{\mu}\tilde{u}^{\nu}~\bar{\chi} s_{103}^{\epsilon} \otimes \gamma_{\nu} \gamma_{\epsilon} \chi, \qquad
T_{104}^{\mu} = d^{\nu}\tilde{u}^{\mu}~\bar{\chi} s_{104}^{\epsilon} \otimes \gamma_{\nu} \gamma_{\epsilon} \chi,
\nonumber\\
T_{105}^{\mu} = d_{\nu}\tilde{u}^{\nu}~\bar{\chi} s_{105}^{\epsilon} \otimes \gamma^{\mu} \gamma_{\epsilon} \chi, \qquad
T_{106}^{\mu} = u^{\nu}~\bar{\tilde{\chi}} s_{106}^{\epsilon} \otimes \gamma^{\mu} \gamma_{\epsilon} d_{\nu}\chi,
\nonumber\\
T_{107}^{\mu} = u^{\nu}~\bar{\tilde{\chi}} s_{107}^{\epsilon} \otimes \gamma_{\nu} \gamma_{\epsilon} d^{\mu}\chi, \qquad
T_{108}^{\mu} = u^{\nu}~d^{\mu}\bar{\tilde{\chi}} s_{108}^{\epsilon} \otimes \gamma_{\nu} \gamma_{\epsilon} \chi,
\nonumber\\
T_{109}^{\mu} = u^{\nu}~d_{\nu}\bar{\tilde{\chi}} s_{109}^{\epsilon} \otimes \gamma^{\mu} \gamma_{\epsilon} \chi, \qquad
T_{110}^{\mu} = d^{\mu}u^{\nu}~\bar{\tilde{\chi}} s_{110}^{\epsilon} \otimes \gamma_{\nu} \gamma_{\epsilon} \chi,
\nonumber\\
T_{111}^{\mu} = d^{\nu}u^{\mu}~\bar{\tilde{\chi}} s_{111}^{\epsilon} \otimes \gamma_{\nu} \gamma_{\epsilon} \chi, \qquad
T_{112}^{\mu} = d_{\nu}u^{\nu}~\bar{\tilde{\chi}} s_{112}^{\epsilon} \otimes \gamma^{\mu} \gamma_{\epsilon} \chi,
\nonumber\\
T_{113}^{\mu} = u^{\nu}~\bar{\chi} s_{113}^{\epsilon} \otimes \gamma^{\mu} \gamma_{\epsilon} d_{\nu}\tilde{\chi}, \qquad
T_{114}^{\mu} = u^{\nu}~\bar{\chi} s_{114}^{\epsilon} \otimes \gamma_{\nu} \gamma_{\epsilon} d^{\mu}\tilde{\chi},
\nonumber\\
T_{115}^{\mu} = u^{\nu}~d^{\mu}\bar{\chi} s_{115}^{\epsilon} \otimes \gamma_{\nu} \gamma_{\epsilon} \tilde{\chi}, \qquad
T_{116}^{\mu} = u^{\nu}~d_{\nu}\bar{\chi} s_{116}^{\epsilon} \otimes \gamma^{\mu} \gamma_{\epsilon} \tilde{\chi},
\nonumber\\
T_{117}^{\mu} = d^{\mu}u^{\nu}~\bar{\chi} s_{117}^{\epsilon} \otimes \gamma_{\nu} \gamma_{\epsilon} \tilde{\chi}, \qquad
T_{118}^{\mu} = d^{\nu}u^{\mu}~\bar{\chi} s_{118}^{\epsilon} \otimes \gamma_{\nu} \gamma_{\epsilon} \tilde{\chi},
\nonumber\\
T_{119}^{\mu} = d_{\nu}u^{\nu}~\bar{\chi} s_{119}^{\epsilon} \otimes \gamma^{\mu} \gamma_{\epsilon} \tilde{\chi}.
\eea
We substitute these expressions in (\ref{gauge1a}) and we obtain the solutions from the statment. 
$\qed$
\newpage

\section{Second Order Gauge Invariance\label{second}}

We proceed now to the second order analysis as in \cite{wick+hopf}. First we give the expressions of the Wick submonomials.
Some of the expressions from (\ref{sub1}) get some supplementary contributions from the spin $3/2$ field:
\bea
C_{a\mu} \equiv v_{a\mu} \cdot T = f_{abc} (v_{b}^{\nu}~F_{c\nu\mu} - u_{b}~\tilde{u}_{c,\mu}) 
+ \bar{\Psi}_{\nu}~t_{a}^{\epsilon} \otimes \gamma^{\nu}\gamma_{\mu}\gamma^{\rho}\gamma_{\epsilon}\Psi_{\rho}
\nonumber\\
D_{a} \equiv u_{a} \cdot T = f_{abc} v^{\mu}_{b}~\tilde{u}_{c,\mu} +
2 ( \bar{\Psi}_{\mu}~t_{a}^{\epsilon} \otimes \gamma^{\mu}\gamma_{\epsilon}\tilde{\chi}
- \bar{\tilde{\chi}}~t_{a}^{\epsilon} \otimes \gamma^{\mu}\gamma_{\epsilon}\Psi_{\mu} ).
\label{sub1a}
\eea
and we also have:
\bea
\bar{F}^{\mu}_{A\alpha} \equiv \Psi^{\mu}_{A\alpha}\cdot T = 
- v_{a}^{\rho}~\bar{\Psi}^{\nu}_{B\beta} (\gamma_{\nu}\gamma_{\rho}\gamma^{\mu}\gamma_{\epsilon})_{\beta\alpha} (t_{a}^{\epsilon})_{BA} 
+ 2 u_{a}~\bar{\tilde{\chi}}_{B\beta} (\gamma^{\mu}\gamma_{\epsilon})_{\beta\alpha} (t_{a}^{\epsilon})_{BA}
\nonumber\\
F^{\mu}_{A\alpha} \equiv \bar{\Psi}^{\mu}_{A\alpha}\cdot T = 
v_{a}^{\rho}~(t_{a}^{\epsilon})_{AB}~(\gamma^{\mu}\gamma_{\rho}\gamma_{\nu}\gamma_{\epsilon})_{\alpha\beta} \Psi^{\nu}_{B\beta}
- 2 u_{a}~(t_{a}^{\epsilon})_{AB}~(\gamma^{\mu}\gamma_{\epsilon})_{\alpha\beta} \tilde{\chi}_{B\beta} 
\nonumber\\
\bar{F}^{\mu\nu}_{A\alpha} \equiv \Psi^{\mu}_{A\alpha}\cdot T^{\nu} = 
- u_{a}~\bar{\Psi}^{\rho}_{B\beta} (\gamma_{\rho}\gamma^{\nu}\gamma^{\mu}\gamma_{\epsilon})_{\beta\alpha}
(t_{a}^{\epsilon})_{BA} 
\nonumber\\
F^{\mu\nu}_{A\alpha} \equiv \bar{\Psi}^{\mu}_{A\alpha}\cdot T^{\nu} = 
u_{a}~(t_{a}^{\epsilon})_{AB}~(\gamma^{\mu}\gamma^{\nu}\gamma_{\rho}\gamma_{\epsilon})_{\alpha\beta} \Psi^{\rho}_{B\beta}
\nonumber\\
\bar{G}_{A\alpha} \equiv \tilde{\chi}_{A\alpha}\cdot T = 
2 u_{a}~\Psi^{\mu}_{B\beta} (\gamma_{\mu}\gamma_{\epsilon})_{\beta\alpha} (t_{a}^{\epsilon})_{BA} 
\nonumber\\
G_{A\alpha} \equiv \bar{\tilde{\chi}}_{A\alpha}\cdot T = 
- 2 u_{a}~(t_{a}^{\epsilon})_{AB}~(\gamma_{\mu}\gamma_{\epsilon})_{\alpha\beta} \Psi^{\mu}_{B\beta}
\eea
or in matrix notation
\bea
\bar{F}^{\mu} = - v_{a}^{\rho}~\bar{\Psi}^{\nu} t_{a}^{\epsilon} \otimes \gamma_{\nu}\gamma_{\rho}\gamma^{\mu}\gamma_{\epsilon} 
+ 2 u_{a}~\bar{\tilde{\chi}} t_{a}^{\epsilon} \otimes \gamma^{\mu}\gamma_{\epsilon}
\nonumber\\
F^{\mu} = v_{a}^{\rho}~t_{a}^{\epsilon} \otimes \gamma^{\mu}\gamma_{\rho}\gamma_{\nu}\gamma_{\epsilon} \Psi^{\nu}
- 2 u_{a}~t_{a}^{\epsilon} \otimes \gamma^{\mu}\gamma_{\epsilon} \tilde{\chi} 
\nonumber\\
\bar{F}^{\mu\nu} = - u_{a}~\bar{\Psi}^{\rho} t_{a}^{\epsilon} \otimes \gamma_{\rho}\gamma^{\nu}\gamma^{\mu}\gamma_{\epsilon} 
\nonumber\\
F^{\mu\nu} = u_{a}~t_{a}^{\epsilon} \otimes \gamma^{\mu}\gamma^{\nu}\gamma_{\rho}\gamma_{\epsilon} \Psi^{\rho}
\nonumber\\
\bar{G} = 2 u_{a}~\Psi^{\mu} t_{a}^{\epsilon} \otimes \gamma_{\mu}\gamma_{\epsilon} 
\nonumber\\
G = - 2 u_{a}~t_{a}^{\epsilon} \otimes \gamma_{\mu}\gamma_{\epsilon} \Psi^{\mu}
\label{psi}
\eea

The relations
\be
s C_{a\mu} \sim 0, \qquad sD_{a} \sim 0
\ee
stay true and we also have
\be
s F^{\mu} \sim 0, \qquad s\bar{F}^{\mu} \sim 0, \qquad sG \sim 0, \qquad s\bar{G} \sim 0.
\ee

Now we can compute the second order anomalies as in \cite{wick+hopf}. For the one-loop contributions we have
\bea
T(T^{(1)}(x_{1}), T^{(1)}(x_{2})) = \cdots + \bar{\Psi}_{\mu A\alpha}(x_{1}) \Psi_{\nu B\beta}(x_{2})~
T(F_{A\alpha}^{\mu(0)}(x_{1}), \bar{F}_{B\beta}^{\nu(0)}(x_{2})) + (x_{1} \leftrightarrow x_{2})
\nonumber\\
= \cdots - 2 i~\partial_{\mu}d^{F}_{0,0}(x_{1} - x_{2})~
\bar{\Psi}_{\mu}(x_{1}) t_{a}^{\epsilon} t_{b}^{\epsilon} \otimes 
\gamma^{\mu} \gamma^{\rho} \gamma^{\nu} \Psi_{\nu}(x_{2}) + (x_{1} \leftrightarrow x_{2})
\eea
where 
$\cdots$
is the pure Yang-Mills contribution. From here we obtain that gauge invariance remains true at one-loop level: 
\be
sT(T^{I(1)}(x_{1}), T^{J(1)}(x_{2})) = 0.
\ee

For tree contributions we have
\bea
T(T^{(2)}(x_{1}), T^{(2)}(x_{2})) = \cdots - i~\bar{F}^{\mu}(x_{1})~S^{F}_{0}(x_{1} - x_{2})~F_{\mu}(x_{2})
+ (x_{1} \leftrightarrow x_{2})
\nonumber\\
T(T^{\mu(2)}(x_{1}), T^{(2)}(x_{2})) = \cdots - i~\bar{F}^{\nu\mu}(x_{1})~S^{F}_{0}(x_{1} - x_{2})~F_{\nu}(x_{2})
\nonumber\\
+ i~\bar{F}_{\nu}(x_{1})~S^{F}_{0}(x_{2} - x_{1})~F^{\nu\mu}(x_{2})
\nonumber\\
T(T^{\mu(2)}(x_{1}), T^{\nu(2)}(x_{2})) = \cdots 
+ i~\bar{F}^{\rho\mu}(x_{1})~S^{F}_{0}(x_{1} - x_{2})~{F_{\rho}}^{\nu}(x_{2})
- (x_{1} \leftrightarrow x_{2}, \mu \leftrightarrow \nu)
\eea
and the gauge invariance equations
\be
sT(T^{I(2)}(x_{1}), T^{J(2)}(x_{2})) = 0
\ee
are true {\it iff}
\be
[ t_{a}^{\epsilon}, t_{b}^{\epsilon} ] = i~f_{abc}~t_{c}^{\epsilon}.
\ee

We can do the same analysis for the coupling between gravity and particles of spin $3/2$. Suprisingly, we get a negative result.
The computations leading to the anomaly 
\be
A = sT(T^{(2)}(x_{1}), T^{(2)}(x_{2}))
\ee
are extremly long. However, to obtain the negative result anticipated above, we can proceed as follows. First we consider the 
piece of the anomaly 
$
A \sim \bar{\chi} \chi
$
which is:
\bea
A_{\chi} = - 2 i~d_{\rho}h \tilde{u}_{\mu} d^{\rho}\bar{\chi} (t_{2}^{\epsilon})^{2} \otimes \gamma^{\mu}\gamma_{\epsilon}\chi 
+ 2 i~d^{\lambda}h \tilde{u}_{\mu} d^{\mu}\bar{\chi} (t_{2}^{\epsilon})^{2} \otimes \gamma_{\lambda}\gamma_{\epsilon}\chi 
\nonumber\\
- 2 i~h \tilde{u}_{\mu} d^{\rho}\bar{\chi} (t_{2}^{\epsilon})^{2} \otimes \gamma^{\mu}\gamma_{\epsilon}d_{\rho}\chi 
+ 2 i~d^{\lambda}h_{\rho\sigma} \tilde{u}_{\mu} 
d^{\rho}\bar{\chi} (t_{2}^{\epsilon})^{2} \otimes \gamma^{\mu}\gamma_{\lambda}\gamma^{\sigma}\gamma_{\epsilon}\chi 
\nonumber\\
+ 4 i~h_{\rho\sigma} \tilde{u}_{\mu} d^{\rho}\bar{\chi} (t_{2}^{\epsilon})^{2} \otimes \gamma^{\mu}\gamma_{\epsilon} d^{\sigma}\chi 
+ 2 i~d_{\rho}h \tilde{u}_{\mu} \bar{\chi} (t_{2}^{\epsilon})^{2} \otimes \gamma^{\mu}\gamma_{\epsilon} d^{\rho}\chi 
\nonumber\\
- 2 i~d^{\lambda}h \tilde{u}_{\mu} \bar{\chi} (t_{2}^{\epsilon})^{2} \otimes \gamma_{\lambda}\gamma_{\epsilon} d^{\mu}\chi 
+ 2 i~h \tilde{u}_{\mu} d_{\rho}\bar{\chi} (t_{2}^{\epsilon})^{2} \otimes \gamma^{\mu}\gamma_{\epsilon} d^{\rho}\chi
\nonumber\\
- 4 i~d^{\rho}h_{\rho\sigma} \tilde{u}_{\mu} \bar{\chi} (t_{2}^{\epsilon})^{2} \otimes \gamma^{\mu}\gamma_{\epsilon}d^{\sigma}\chi 
+ 4 i~d^{\nu}h^{\mu\sigma} \tilde{u}_{\mu} \bar{\chi} (t_{2}^{\epsilon})^{2} \otimes \gamma_{\nu}\gamma_{\epsilon}d_{\sigma}\chi 
\nonumber\\
- 4 i~d^{\mu}h^{\rho\sigma} \tilde{u}_{\mu} \bar{\chi} (t_{2}^{\epsilon})^{2} \otimes \gamma_{\rho}\gamma_{\epsilon}d_{\sigma}\chi 
+ 2 i~d_{\nu}h_{\rho\sigma} \tilde{u}_{\mu} \bar{\chi} (t_{2}^{\epsilon})^{2} 
\otimes \gamma^{\mu}\gamma^{\nu}\gamma^{\rho}\gamma_{\epsilon}d^{\sigma}\chi 
\nonumber\\
- 2 i~h_{\rho\sigma} \tilde{u}_{\mu} d^{\sigma}\bar{\chi} (t_{2}^{\epsilon})^{2} \otimes \gamma^{\mu}\gamma_{\epsilon} d^{\rho}\chi 
\eea
Next, we must try to express this anomaly as a coboundary
\be
A_{\chi} = d_{Q}B - i d_{\mu}B^{\mu}.
\ee
The list is rather long:
\be
B = \sum b_{j}~B_{j}
\ee
where:
\bea
B_{1} = \tilde{u}_{\mu}~\tilde{u}_{\nu}~d^{\mu}\bar{\chi}~b_{1}^{\epsilon} \otimes \gamma^{\nu}\gamma_{\epsilon}\chi, \qquad
B_{2} = \tilde{u}_{\mu}~\tilde{u}_{\nu}~\bar{\chi}~b_{2}^{\epsilon} \otimes \gamma^{\nu}\gamma_{\epsilon}d^{\mu}\chi,
\nonumber\\
B_{3} = \tilde{u}_{\mu}~d\cdot \tilde{u}~\bar{\chi}~b_{3}^{\epsilon} \otimes \gamma^{\mu}\gamma_{\epsilon}\chi, \qquad
B_{4} = \tilde{u}_{\mu}~d^{\mu}\tilde{u}^{\nu}~\bar{\chi}~b_{4}^{\epsilon} \otimes \gamma_{\nu}\gamma_{\epsilon}\chi,
\nonumber\\
B_{5} = \tilde{u}_{\mu}~d^{\nu}\tilde{u}^{\mu}~\bar{\chi}~b_{5}^{\epsilon} \otimes \gamma_{\nu}\gamma_{\epsilon}\chi, \qquad
B_{6} = \tilde{u}_{\mu}~d_{\nu}\tilde{u}_{\rho}~\bar{\chi}~b_{6}^{\epsilon} 
\otimes \gamma^{\mu}\gamma_{\nu}\gamma^{\rho}\gamma_{\epsilon}\chi.
\eea
and 
\be
B^{\mu} = \sum b_{j}~B_{j}^{\mu}
\ee
where 
\bea
B_{1}^{\mu} = h^{\mu\nu}~\tilde{u}^{\rho}~d_{\nu}\bar{\chi}~c_{1}^{\epsilon} \otimes \gamma_{\rho}\gamma_{\epsilon}\chi, \qquad
B_{2}^{\mu} = h^{\mu\nu}~\tilde{u}^{\rho}~d_{\rho}\bar{\chi}~c_{2}^{\epsilon} \otimes \gamma_{\nu}\gamma_{\epsilon}\chi,
\nonumber\\
B_{3}^{\mu} = h^{\rho\sigma}~\tilde{u}^{\mu}~d_{\rho}\bar{\chi}~c_{3}^{\epsilon} \otimes \gamma_{\sigma}\gamma_{\epsilon}\chi, \qquad
B_{4}^{\mu} = h_{\nu\rho}~\tilde{u}^{\rho}~d^{\mu}\bar{\chi}~c_{4}^{\epsilon} \otimes \gamma^{\nu}\gamma_{\epsilon}\chi,
\nonumber\\
B_{5}^{\mu} = h_{\nu\rho}~\tilde{u}^{\rho}~d^{\nu}\bar{\chi}~c_{5}^{\epsilon} \otimes \gamma^{\mu}\gamma_{\epsilon}\chi, \qquad
B_{6}^{\mu} = h^{\mu\nu}~\tilde{u}^{\rho}~\bar{\chi}~c_{6}^{\epsilon} \otimes \gamma_{\nu}\gamma_{\epsilon}d_{\rho}\chi
\nonumber\\
B_{7}^{\mu} = h^{\mu\nu}~\tilde{u}^{\rho}~\bar{\chi}~c_{7}^{\epsilon} \otimes \gamma_{\rho}\gamma_{\epsilon}d_{\nu}\chi, \qquad
B_{8}^{\mu} = h^{\rho\sigma}~\tilde{u}^{\mu}~\bar{\chi}~c_{8}^{\epsilon} \otimes \gamma_{\rho}\gamma_{\epsilon}d_{\sigma}\chi,
\nonumber\\
B_{9}^{\mu} = h_{\nu\rho}~\tilde{u}^{\rho}~\bar{\chi}~c_{9}^{\epsilon} \otimes \gamma^{\mu}\gamma_{\epsilon}d^{\nu}\chi, \qquad
B_{10}^{\mu} = h_{\nu\rho}~\tilde{u}^{\rho}~\bar{\chi}~c_{10}^{\epsilon} \otimes \gamma^{\nu}\gamma_{\epsilon}d^{\mu}\chi
\nonumber\\
B_{11}^{\mu} = d^{\mu}h^{\nu\rho}~\tilde{u}_{\rho}~\bar{\chi}~c_{11}^{\epsilon} \otimes \gamma_{\nu}\gamma_{\epsilon}\chi, \qquad
B_{12}^{\mu} = d_{\nu}h^{\mu\nu}~\tilde{u}^{\rho}~\bar{\chi}~c_{12}^{\epsilon} \otimes \gamma_{\rho}\gamma_{\epsilon}\chi,
\nonumber\\
B_{13}^{\mu} = d^{\rho}h^{\mu\nu}~\tilde{u}_{\nu}~\bar{\chi}~c_{13}^{\epsilon} \otimes \gamma_{\rho}\gamma_{\epsilon}\chi, \qquad
B_{14}^{\mu} = d^{\rho}h^{\mu\nu}~\tilde{u}_{\rho}~\bar{\chi}~c_{14}^{\epsilon} \otimes \gamma_{\nu}\gamma_{\epsilon}\chi,
\nonumber\\
B_{15}^{\mu} = d^{\nu}h_{\nu\rho}~\tilde{u}^{\mu}~\bar{\chi}~c_{15}^{\epsilon} \otimes \gamma^{\mu}\gamma_{\epsilon}\chi, \qquad
B_{16}^{\mu} = d^{\nu}h_{\nu\rho}~\tilde{u}^{\rho}~\bar{\chi}~c_{16}^{\epsilon} \otimes \gamma_{\nu}\gamma_{\epsilon}\chi
\nonumber\\
B_{17}^{\mu} = h^{\mu\nu}~d_{\nu}\tilde{u}_{\rho}~\bar{\chi}~c_{17}^{\epsilon} \otimes \gamma^{\rho}\gamma_{\epsilon}\chi, \qquad
B_{18}^{\mu} = h^{\mu\nu}~d_{\rho}\tilde{u}_{\nu}~\bar{\chi}~c_{18}^{\epsilon} \otimes \gamma^{\rho}\gamma_{\epsilon}\chi,
\nonumber\\
B_{19}^{\mu} = h^{\mu\nu}~d\cdot \tilde{u}~\bar{\chi}~c_{19}^{\epsilon} \otimes \gamma_{\nu}\gamma_{\epsilon}\chi, \qquad
B_{20}^{\mu} = h_{\nu\rho}~d^{\mu}\tilde{u}^{\nu}~\bar{\chi}~c_{20}^{\epsilon} \otimes \gamma^{\rho}\gamma_{\epsilon}\chi,
\nonumber\\
B_{21}^{\mu} = h_{\nu\rho}~d^{\nu}\tilde{u}^{\mu}~\bar{\chi}~c_{21}^{\epsilon} \otimes \gamma^{\rho}\gamma_{\epsilon}\chi, \qquad
B_{22}^{\mu} = h_{\nu\rho}~d^{\nu}\tilde{u}^{\rho}~\bar{\chi}~c_{22}^{\epsilon} \otimes \gamma^{\mu}\gamma_{\epsilon}\chi
\nonumber\\
B_{23}^{\mu} = h~\tilde{u}_{\nu}~d^{\mu}\bar{\chi}~c_{23}^{\epsilon} \otimes \gamma^{\nu}\gamma_{\epsilon}\chi, \qquad
B_{24}^{\mu} = h~\tilde{u}_{\nu}~\bar{\chi}~c_{24}^{\epsilon} \otimes \gamma^{\nu}\gamma_{\epsilon}d^{\mu}\chi,
\nonumber\\
B_{25}^{\mu} = h~\tilde{u}_{\nu}~d^{\nu}\bar{\chi}~c_{25}^{\epsilon} \otimes \gamma^{\mu}\gamma_{\epsilon}\chi, \qquad
B_{26}^{\mu} = h~\tilde{u}_{\nu}~\bar{\chi}~c_{26}^{\epsilon} \otimes \gamma^{\mu}\gamma_{\epsilon}d^{\nu}\chi,
\nonumber\\
B_{27}^{\mu} = d^{\mu}h~\tilde{u}_{\nu}~\bar{\chi}~c_{27}^{\epsilon} \otimes \gamma^{\nu}\gamma_{\epsilon}\chi, \qquad
B_{28}^{\mu} = d^{\nu}h~\tilde{u}^{\mu}~\bar{\chi}~c_{28}^{\epsilon} \otimes \gamma^{\nu}\gamma_{\epsilon}\chi,
\nonumber\\
B_{29}^{\mu} = d^{\nu}h~\tilde{u}_{\nu}~\bar{\chi}~c_{29}^{\epsilon} \otimes \gamma^{\mu}\gamma_{\epsilon}\chi, \qquad
B_{30}^{\mu} = h~d^{\mu}\tilde{u}^{\nu}~\bar{\chi}~c_{30}^{\epsilon} \otimes \gamma_{\nu}\gamma_{\epsilon}\chi, 
\nonumber\\
B_{31}^{\mu} = h~d^{\nu}\tilde{u}^{\mu}~\bar{\chi}~c_{31}^{\epsilon} \otimes \gamma_{\nu}\gamma_{\epsilon}\chi, \qquad
B_{32}^{\mu} = h~d\cdot \tilde{u}~\bar{\chi}~c_{32}^{\epsilon} \otimes \gamma^{\mu}\gamma_{\epsilon}\chi
\nonumber\\
B_{33}^{\mu} = h_{\nu\rho}~\tilde{u}_{\sigma}~d^{\nu}\bar{\chi}~c_{33}^{\epsilon} 
\otimes \gamma^{\sigma}\gamma^{\mu}\gamma_{\rho}\gamma_{\epsilon}\chi, \qquad
B_{34}^{\mu} = h_{\nu\rho}~\tilde{u}_{\sigma}~\bar{\chi}~c_{34}^{\epsilon} 
\otimes \gamma^{\sigma}\gamma^{\mu}\gamma^{\rho}\gamma_{\epsilon}d^{\nu}\chi,
\nonumber\\
B_{35}^{\mu} = d^{\rho}h^{\mu\nu}~\tilde{u}^{\sigma}~\bar{\chi}~c_{35}^{\epsilon} 
\otimes \gamma_{\sigma}\gamma_{\rho}\gamma_{\nu}\gamma_{\epsilon}\chi, \qquad
B_{36}^{\mu} = d^{\nu}h_{\nu\rho}~\tilde{u}_{\sigma}~\bar{\chi}~c_{36}^{\epsilon} 
\otimes \gamma^{\sigma}\gamma^{\mu}\gamma^{\rho}\gamma_{\epsilon}\chi,
\nonumber\\
B_{37}^{\mu} = h^{\mu\nu}~d^{\rho}\tilde{u}^{\sigma}~\bar{\chi}~c_{37}^{\epsilon} 
\otimes \gamma_{\nu}\gamma_{\rho}\gamma_{\sigma}\gamma_{\epsilon}\chi, \qquad
B_{38}^{\mu} = h^{\rho\sigma}~d_{\rho}\tilde{u}_{\nu}~\bar{\chi}~c_{38}^{\epsilon} 
\otimes \gamma_{\sigma}\gamma^{\mu}\gamma^{\nu}\gamma_{\epsilon}\chi
\eea
Now we compute
\bea
A_{\chi} - sB \equiv A_{\chi} - (d_{Q}B - i d_{\mu}B^{\mu}) =
\nonumber\\
i~[\tilde{u}_{\mu}~d^{\rho}h^{\mu\nu}~d_{\nu}\bar{\chi}~(c_{5}^{\epsilon} + c_{13}^{\epsilon}) \otimes \gamma_{\rho}\gamma_{\epsilon}\chi
+ \tilde{u}_{\mu}~d_{\nu}h~d^{\mu}\bar{\chi}~(2 t^{\epsilon} + c_{25}^{\epsilon} + c_{28}^{\epsilon}) 
\otimes \gamma^{\nu}\gamma_{\epsilon}\chi
\nonumber\\
+ \tilde{u}_{\mu}~d^{\rho}h^{\mu\nu}~\bar{\chi}~(4 t^{\epsilon} + c_{9}^{\epsilon} + c_{23}^{\epsilon})
\otimes \gamma_{\rho}\gamma_{\epsilon}d_{\nu}\chi
+ \tilde{u}_{\mu}~d_{\nu}h~\bar{\chi}~(- 2 t^{\epsilon} + c_{26}^{\epsilon} + c_{28}^{\epsilon})
\otimes \gamma^{\nu}\gamma_{\epsilon}d^{\mu}\chi
\nonumber\\
+ d_{\mu}\tilde{u}_{\nu}~d_{\rho}h^{\nu\rho}~
\bar{\chi}~(- b_{5}^{\epsilon} + c_{16}^{\epsilon} + c_{18}^{\epsilon} - 2 c_{36}^{\epsilon}) 
\otimes \gamma^{\mu}\gamma_{\epsilon}\chi
+ d_{\mu}\tilde{u}_{\nu}~d^{\rho}h^{\mu\nu}~\bar{\chi}~(c_{13}^{\epsilon} + c_{22}^{\epsilon}) 
\otimes \gamma_{\rho}\gamma_{\epsilon}\chi
\nonumber\\
+ d_{\mu}\tilde{u}_{\nu}~d^{\nu}h~\bar{\chi}~(c_{29}^{\epsilon} + c_{31}^{\epsilon}) \otimes \gamma^{\mu}\gamma_{\epsilon}\chi
+ d\cdot \tilde{u}~d_{\rho}h~\bar{\chi}~(c_{28}^{\epsilon} + c_{32}^{\epsilon}) \otimes \gamma^{\rho}\gamma_{\epsilon}\chi
\nonumber\\
+ d_{\mu}\tilde{u}_{\nu}~h^{\nu\rho}~
d_{\rho}\bar{\chi}~(c_{5}^{\epsilon} + c_{18}^{\epsilon} - 2 c_{37}^{\epsilon}) \otimes \gamma^{\mu}\gamma_{\epsilon}\chi
\nonumber\\
+ d_{\mu} \tilde{u}_{\nu}~h^{\nu\rho}~
\bar{\chi}~(c_{9}^{\epsilon} + c_{18}^{\epsilon} - 2 c_{37}^{\epsilon}) \otimes \gamma^{\mu}\gamma_{\epsilon}d_{\rho}\chi ] + \cdots
\eea
where 
$
t^{\epsilon} \equiv (t_{2}^{\epsilon})^{2}
$
and
$\cdots$ are other (linear independent) contributions. If we equate to $0$ the right hand side (this is a necessary for the 
elimination of the anomaly by redefinitions of the chronological products) in the end we obtain
$
(t_{2}^{\epsilon})^{2} = 0 \quad \Rightarrow \quad t_{2}^{\epsilon} = 0
$
because
$
t_{2}^{\epsilon}
$
is self-adjoint.
\newpage
The next step is to compute the anomaly using this simplification. A tedious computation leads (up to the equations of motion) to:
\bea
A = i~h_{\mu\nu}~u_{\lambda}~[ 20 ( d^{\lambda}\bar{\Psi}_{\rho} t^{\epsilon} 
\otimes \gamma^{\rho}\gamma^{\mu}\gamma^{\sigma}\gamma_{\epsilon} d^{\nu}\Psi_{\sigma} 
- d^{\nu}\bar{\Psi}_{\rho} t^{\epsilon} 
\otimes \gamma^{\rho}\gamma^{\mu}\gamma^{\sigma}\gamma_{\epsilon} d^{\lambda}\Psi_{\sigma})
\nonumber\\
+ 8 ( d^{\nu}\bar{\Psi}_{\rho} t^{\epsilon} 
\otimes \gamma^{\rho}\gamma^{\mu}\gamma^{\lambda}\gamma_{\epsilon} d_{\sigma}\Psi^{\sigma} 
- d_{\rho}\bar{\Psi}^{\rho} t^{\epsilon} 
\otimes \gamma^{\lambda}\gamma^{\mu}\gamma^{\sigma}\gamma_{\epsilon} d^{\nu}\Psi_{\sigma}) ]
\nonumber\\
+ 4 i~h_{\mu\nu}~d_{\alpha}u_{\beta}~( \bar{\Psi}_{\rho} t^{\epsilon} 
\otimes \gamma^{\rho}\gamma^{\beta}\gamma^{\alpha}\gamma^{\mu}\gamma^{\sigma}\gamma_{\epsilon} d^{\nu}\Psi_{\sigma} 
- d^{\nu}\bar{\Psi}_{\rho} t^{\epsilon} 
\otimes \gamma^{\rho}\gamma^{\mu}\gamma^{\beta}\gamma^{\alpha}\gamma^{\sigma}\gamma_{\epsilon} \Psi_{\sigma})
\nonumber\\
+ i~u_{\mu}~u_{\nu}[ 20 ( d^{\mu}\bar{\tilde{\chi}} t^{\epsilon} \otimes \gamma_{\rho}\gamma_{\epsilon} d^{\nu}\Psi^{\rho} 
- d^{\mu}\bar{\Psi}^{\rho} t^{\epsilon} \otimes \gamma_{\rho}\gamma_{\epsilon} d^{\nu}\tilde{\chi})
\nonumber\\
+ 8 ( d_{\rho}\bar{\Psi}^{\rho} t^{\epsilon} \otimes \gamma^{\mu}\gamma_{\epsilon} d^{\nu}\tilde{\chi} 
- d^{\mu}\bar{\tilde{\chi}} t^{\epsilon} \otimes \gamma^{\nu}\gamma_{\epsilon} d_{\rho}\Psi^{\rho}) ]
\nonumber\\
+ 4 i~u_{\mu}~d_{\alpha}u_{\beta} 
( d^{\mu}\bar{\tilde{\chi}} t^{\epsilon} \otimes \gamma^{\alpha}\gamma^{\beta}\gamma^{\rho}\gamma_{\epsilon} \Psi_{\rho} 
- \bar{\Psi}_{\rho} t^{\epsilon} \otimes \gamma^{\rho}\gamma^{\beta}\gamma^{\alpha}\gamma_{\epsilon} d^{\mu}\tilde{\chi})
\nonumber\\
- 20 i~u_{\mu}~d\cdot u 
( \bar{\tilde{\chi}} t^{\epsilon} \otimes \gamma_{\rho}\gamma_{\epsilon} d^{\mu}\Psi^{\rho} 
+ d^{\mu}\bar{\Psi}^{\rho} t^{\epsilon} \otimes \gamma_{\rho}\gamma_{\epsilon} d^{\nu}\tilde{\chi})
\nonumber\\
+ 4 i~d\cdot u~d_{\alpha} u_{\beta} 
( \bar{\tilde{\chi}} t^{\epsilon} \otimes \gamma^{\alpha}\gamma^{\beta}\gamma^{\rho}\gamma_{\epsilon} \Psi_{\rho} 
- \bar{\Psi}_{\rho} t^{\epsilon} \otimes \gamma^{\rho}\gamma^{\beta}\gamma^{\alpha}\gamma_{\epsilon} \tilde{\chi})
+ d_{Q}B
\label{A}
\eea
where the last term is a coboundary which we do not write explicitly and now
$
t^{\epsilon} \equiv (t^{\epsilon})^{2}.
$

We have to rewrite strings gamma metrices using an independent basis. We remind this trick. 
For two gamma matrices, this is elementary: we define
\be
\Sigma^{\mu\nu} \equiv {1\over 2}~(\gamma^{\mu} \gamma^{\nu} - \gamma^{\nu} \gamma^{\mu})
\ee
and we have
\be
\gamma^{\mu} \gamma^{\nu} = \Sigma^{\mu\nu} + \eta^{\mu\nu} {\bf 1}.
\ee
For three gamma matrices the independent basis is given by the totally antisymmetric expressions:
\be
\Sigma^{\mu\nu\rho} \equiv {1\over 6} (\gamma^{\mu} \gamma^{\nu} \gamma^{\rho} 
+ \gamma^{\nu} \gamma^{\rho} \gamma^{\mu} + \gamma^{\rho} \gamma^{\mu} \gamma^{\nu}
- \gamma^{\nu} \gamma^{\mu} \gamma^{\rho} - \gamma^{\mu} \gamma^{\rho} \gamma^{\nu} - \gamma^{\rho} \gamma^{\nu} \gamma^{\mu})   
\ee
and we can easily express 
$
\gamma^{\mu} \gamma^{\nu} \gamma^{\rho} 
$
in terms of independent matrices:
\be
\gamma^{\mu} \gamma^{\nu} \gamma^{\rho} = \Sigma^{\mu\nu\rho} 
- \eta^{\mu\rho}~\gamma^{\nu} + \eta^{\mu\nu}~\gamma^{\rho} + \eta^{\nu\rho}~\gamma^{\mu}.
\ee

Now we go to strings of four gamma matrices and define the totally antisymmetric expressions:
\be
\Sigma^{\mu\nu\alpha\beta} \equiv {1\over 4} (\Sigma^{\mu\nu\alpha} \gamma^{\beta} - \Sigma^{\nu\alpha\beta} \gamma^{\mu}
+ \Sigma^{\mu\alpha\beta} \gamma^{\nu} - \Sigma^{\mu\nu\beta} \gamma^{\alpha})
\ee
and can express 
$
\gamma^{\mu} \gamma^{\nu} \gamma^{\alpha} \gamma^{\beta} 
$
in terms of independent matrices:
\bea
\gamma^{\mu} \gamma^{\nu} \gamma^{\alpha} \gamma^{\beta}  = \Sigma^{\mu\nu\alpha\beta} 
- \eta^{\mu\alpha}~\Sigma^{\nu\beta} + \eta^{\mu\nu}~\Sigma^{\alpha\beta} + \eta^{\nu\alpha}~\Sigma^{\mu\beta}
+ \eta^{\mu\beta}~\Sigma^{\nu\alpha} - \eta^{\nu\beta}~\Sigma^{\mu\alpha} + \eta^{\alpha\beta}~\Sigma^{\mu\nu}
\nonumber\\
+ ( - \eta^{\mu\alpha}~\eta^{\nu\beta} + \eta^{\nu\alpha}~\eta^{\mu\beta} + \eta^{\mu\nu}~\eta^{\alpha\beta}) {\bf 1}.
\eea
To obtain strings of five gamma matrices we define by analogy the totally antisymmetric expressions:
\be
\Sigma^{\mu\nu\rho\alpha\beta} \equiv {1\over 5} (\Sigma^{\mu\nu\rho\alpha} \gamma^{\beta} - \Sigma^{\mu\nu\rho\beta} \gamma^{\alpha}
+ \Sigma^{\mu\nu\alpha\beta} \gamma^{\rho} - \Sigma^{\mu\rho\alpha\beta} \gamma^{\nu} + \Sigma^{\nu\rho\alpha\beta} \gamma^{\mu}).
\ee
But in four dimensions such an expression must be null, so we have the identity:
\be
\Sigma^{\mu\nu\rho\alpha} \gamma^{\beta} - \Sigma^{\mu\nu\rho\beta} \gamma^{\alpha}
+ \Sigma^{\mu\nu\alpha\beta} \gamma^{\rho} - \Sigma^{\mu\rho\alpha\beta} \gamma^{\nu} 
+ \Sigma^{\nu\rho\alpha\beta} \gamma^{\mu} = 0.
\ee
If we use the preceding formulas we obtain
\bea
\gamma^{\mu} \gamma^{\nu} \gamma^{\rho} \gamma^{\alpha} \gamma^{\beta}  =
\eta^{\mu\nu}~\Sigma^{\rho\alpha\beta} - \eta^{\mu\rho}~\Sigma^{\nu\alpha\beta} + \eta^{\mu\alpha}~\Sigma^{\nu\rho\beta}
- \eta^{\mu\beta}~\Sigma^{\nu\rho\alpha}
\nonumber\\
+ \eta^{\nu\rho}~\Sigma^{\mu\alpha\beta} - \eta^{\nu\alpha}~\Sigma^{\mu\rho\beta} + \eta^{\nu\beta}~\Sigma^{\mu\rho\alpha}
+ \eta^{\rho\alpha}~\Sigma^{\mu\rho\beta} - \eta^{\rho\beta}~\Sigma^{\mu\nu\alpha} + \eta^{\alpha\beta}~\Sigma^{\mu\nu\rho}
\nonumber\\
- ( \eta^{\nu\rho} \eta^{\alpha\beta} + \eta^{\rho\alpha} \eta^{\nu\beta} - \eta^{\nu\alpha} \eta^{\rho\beta}) \gamma^{\mu}
+ ( \eta^{\mu\rho} \eta^{\alpha\beta} + \eta^{\rho\alpha} \eta^{\mu\beta} - \eta^{\mu\alpha} \eta^{\rho\beta}) \gamma^{\nu}
\nonumber\\
- ( \eta^{\mu\nu} \eta^{\alpha\beta} + \eta^{\nu\alpha} \eta^{\mu\beta} - \eta^{\mu\alpha} \eta^{\nu\beta}) \gamma^{\rho}
+ ( \eta^{\mu\nu} \eta^{\rho\beta} + \eta^{\nu\rho} \eta^{\mu\beta} - \eta^{\mu\rho} \eta^{\nu\beta}) \gamma^{\alpha}
\nonumber\\
- ( \eta^{\mu\nu} \eta^{\rho\alpha} + \eta^{\nu\rho} \eta^{\mu\alpha} - \eta^{\mu\rho} \eta^{\nu\alpha}) \gamma^{\beta}.
\eea

Now we consider the contributions 
$
A_{hu} \sim h_{\mu\nu} u_{\lambda} \bar{\Psi}_{\alpha} \Psi_{\beta}
$
of the anomaly (\ref{A}) and impose the condition
\be
A_{hu} = sa \equiv d_{Q}a - i d_{\mu}b^{\mu}
\ee
necessary for the elimination of the anomaly by redefinitions of the chronological products. We give the generic expressions
\be
a = \sum a_{j}, \qquad b^{\mu} = \sum b_{j}^{\mu}.
\ee
\bea
a_{1} = h_{\mu\nu}~u_{\lambda}~d^{\nu}\bar{\Psi}_{\rho} a_{1}^{\epsilon} 
\otimes \Sigma^{\mu\lambda\rho} \gamma_{\epsilon} \tilde{\chi}, \qquad
a_{2} = h_{\mu\nu}~u_{\lambda}~\bar{\tilde{\chi}} a_{2}^{\epsilon} 
\otimes \Sigma^{\mu\lambda\rho} \gamma_{\epsilon} d^{\nu}\Psi_{\rho},
\nonumber\\
a_{3} = h_{\mu\nu}~h_{\alpha\beta}~d^{\nu}\bar{\Psi}^{\alpha} a_{3}^{\epsilon} 
\otimes \Sigma^{\mu\beta\lambda} \gamma_{\epsilon} \Psi_{\lambda}, \qquad
a_{4} = h_{\alpha\nu}~{h_{\beta}}^{\nu}~d^{\alpha}\bar{\Psi}_{\rho} a_{4}^{\epsilon} 
\otimes \Sigma^{\beta\rho\lambda} \gamma_{\epsilon} \Psi_{\lambda},
\nonumber\\
a_{5} = h_{\mu\nu}~h~d^{\nu}\bar{\Psi}_{\rho} a_{5}^{\epsilon} 
\otimes \Sigma^{\mu\rho\lambda} \gamma_{\epsilon} \Psi_{\lambda}, \qquad
a_{6} = h_{\alpha\nu}~h_{\alpha\beta}~d^{\nu}\bar{\Psi}_{\rho} a_{6}^{\epsilon} 
\otimes \Sigma^{\mu\alpha\rho} \gamma_{\epsilon} \Psi^{\beta},
\nonumber\\
a_{7} = h_{\mu\nu}~h_{\alpha\beta}~\bar{\Psi}_{\alpha} a_{7}^{\epsilon} 
\otimes \Sigma^{\beta\mu\lambda} \gamma_{\epsilon} d^{\nu}\Psi_{\lambda}, \qquad
a_{8} = h_{\alpha\nu}~h~\bar{\Psi}_{\lambda} a_{8}^{\epsilon} 
\otimes \Sigma^{\mu\lambda\rho} \gamma_{\epsilon} d^{\nu}\Psi_{\rho},
\nonumber\\
a_{9} = h_{\alpha\nu}~{h_{\beta}}^{\nu}~\bar{\Psi}^{\lambda} a_{9}^{\epsilon} 
\otimes \Sigma^{\alpha\lambda\rho} \gamma_{\epsilon} d^{\beta}\Psi_{\rho}, \qquad
a_{10} = h_{\mu\nu}~h_{\alpha\beta}~\bar{\Psi}_{\lambda} a_{10}^{\epsilon} 
\otimes \Sigma^{\mu\alpha\lambda} \gamma_{\epsilon} d^{\nu}\Psi^{\beta},
\nonumber\\
a_{11} = h~d_{\alpha}u_{\beta}~\bar{\tilde{\chi}} a_{11}^{\epsilon} 
\otimes \Sigma^{\alpha\beta\rho} \gamma_{\epsilon} \Psi_{\rho}, \qquad
a_{12} = h~d_{\alpha}u_{\beta}~\bar{\Psi}_{\rho} a_{12}^{\epsilon} 
\otimes \Sigma^{\alpha\beta\rho} \gamma_{\epsilon} \tilde{\chi},
\nonumber\\
a_{13} = h_{\mu\nu}~u_{\lambda}~\bar{\Psi}_{\rho} a_{13}^{\epsilon} 
\otimes \Sigma^{\mu\lambda\rho} \gamma_{\epsilon} d^{\nu}\tilde{\chi}, \qquad
a_{14} = h_{\mu\nu}~u_{\lambda}~d^{\mu}\bar{\tilde{\chi}} a_{14}^{\epsilon} 
\otimes \Sigma^{\nu\lambda\rho} \gamma_{\epsilon} \Psi_{\rho},
\nonumber\\
a_{15} = h_{\mu\lambda}~d^{\lambda}u_{\nu}~\bar{\tilde{\chi}} a_{15}^{\epsilon} 
\otimes \Sigma^{\mu\nu\rho} \gamma_{\epsilon} \Psi_{\rho}, \qquad
a_{16} = h_{\mu\lambda}~d^{\lambda}u_{\nu}~\bar{\Psi}_{\rho} a_{16}^{\epsilon} 
\otimes \Sigma^{\mu\nu\rho} \gamma_{\epsilon} \tilde{\chi},
\nonumber\\
a_{17} = u_{\alpha}~\tilde{u}_{\beta}~d_{\rho}\bar{\Psi}_{\nu} a_{17}^{\epsilon} 
\otimes \Sigma^{\alpha\beta\nu} \gamma_{\epsilon} \Psi^{\rho}, \qquad
a_{18} = u_{\alpha}~\tilde{u}_{\beta}~\bar{\Psi}_{\rho} a_{18}^{\epsilon} 
\otimes \Sigma^{\alpha\beta\nu} \gamma_{\epsilon} d_{\rho}\Psi_{\nu},
\nonumber\\
a_{19} = u_{\mu}~d_{\nu}h~\bar{\tilde{\chi}} a_{19}^{\epsilon} 
\otimes \Sigma^{\mu\nu\rho} \gamma_{\epsilon} \Psi^{\rho}, \qquad
a_{20} = u_{\mu}~d_{\nu}h~\bar{\Psi}_{\rho} a_{20}^{\epsilon} 
\otimes \Sigma^{\mu\nu\rho} \gamma_{\epsilon} \tilde{\chi},
\nonumber\\
a_{21} = d_{\alpha}u_{\beta}~\tilde{u}_{\mu}~\bar{\Psi}^{\mu} a_{21}^{\epsilon} 
\otimes \Sigma^{\alpha\beta\rho} \gamma_{\epsilon} \Psi_{\rho}, \qquad
a_{22} = d_{\alpha}u_{\beta}~\tilde{u}_{\mu}~\bar{\Psi}^{\alpha} a_{22}^{\epsilon} 
\otimes \Sigma^{\mu\beta\rho} \gamma_{\epsilon} \Psi_{\rho},
\nonumber\\
a_{23} = d_{\alpha}u_{\beta}~\tilde{u}_{\mu}~\bar{\Psi}^{\beta} a_{23}^{\epsilon} 
\otimes \Sigma^{\mu\alpha\rho} \gamma_{\epsilon} \Psi_{\rho}, \qquad
a_{24} = d_{\alpha}u_{\beta}~\tilde{u}_{\mu}~\bar{\Psi}_{\rho} a_{24}^{\epsilon} 
\otimes \Sigma^{\mu\alpha\beta} \gamma_{\epsilon} \Psi^{\rho},
\nonumber\\
a_{25} = d_{\alpha}u_{\beta}~\tilde{u}_{\mu}~\bar{\Psi}_{\rho} a_{25}^{\epsilon} 
\otimes \Sigma^{\alpha\beta\rho} \gamma_{\epsilon} \Psi^{\mu}, \qquad
a_{26} = d_{\alpha}u_{\beta}~\tilde{u}_{\mu}~\bar{\Psi}_{\rho} a_{26}^{\epsilon} 
\otimes \Sigma^{\mu\beta\rho} \gamma_{\epsilon} \Psi^{\alpha},
\nonumber\\
a_{27} = d_{\alpha}u_{\beta}~\tilde{u}_{\mu}~\bar{\Psi}_{\rho} a_{27}^{\epsilon} 
\otimes \Sigma^{\mu\alpha\rho} \gamma_{\epsilon} \Psi^{\beta}, \qquad
a_{28} = d^{\nu}h_{\mu\nu}~h_{\alpha\beta}~\bar{\Psi}^{\alpha} a_{28}^{\epsilon} 
\otimes \Sigma^{\mu\beta\rho} \gamma_{\epsilon} \Psi_{\rho},
\nonumber\\
a_{29} = d^{\nu}h_{\mu\nu}~h_{\alpha\beta}~\bar{\Psi}_{\rho} a_{29}^{\epsilon} 
\otimes \Sigma^{\mu\alpha\rho} \gamma_{\epsilon} \Psi^{\beta}, \qquad
a_{30} = d^{\lambda}h_{\mu\nu}~h_{\alpha\lambda}~\bar{\Psi}^{\mu} a_{30}^{\epsilon} 
\otimes \Sigma^{\nu\alpha\rho} \gamma_{\epsilon} \Psi_{\rho},
\nonumber\\
a_{31} = d^{\mu}h~h_{\mu\nu}~\bar{\Psi}^{\rho} a_{31}^{\epsilon} 
\otimes \Sigma^{\nu\rho\lambda} \gamma_{\epsilon} \Psi_{\lambda}, \qquad
a_{32} = d^{\lambda}h_{\mu\nu}~h_{\alpha\lambda}~\bar{\Psi}_{\rho} a_{32}^{\epsilon} 
\otimes \Sigma^{\mu\alpha\rho} \gamma_{\epsilon} \Psi^{\nu},
\nonumber\\
a_{33} = d_{\lambda}h_{\mu\nu}~h^{\nu\lambda}~\bar{\Psi}_{\alpha} a_{33}^{\epsilon} 
\otimes \Sigma^{\mu\alpha\beta} \gamma_{\epsilon} \Psi_{\beta}, \qquad
a_{34} = d_{\lambda}h_{\mu\rho}~{h_{\nu}}^{\rho}~\bar{\Psi}^{\lambda} a_{34}^{\epsilon} 
\otimes \Sigma^{\mu\nu\alpha} \gamma_{\epsilon} \Psi_{\alpha},
\nonumber\\
a_{35} = d_{\lambda}h_{\mu\rho}~{h_{\nu}}^{\rho}~\bar{\Psi}_{\alpha} a_{35}^{\epsilon} 
\otimes \Sigma^{\mu\nu\alpha} \gamma_{\epsilon} \Psi^{\lambda}, \qquad
a_{36} = d_{\lambda}h_{\mu\nu}~{h^{\nu}}_{\alpha}~\bar{\Psi}^{\mu} a_{36}^{\epsilon} 
\otimes \Sigma^{\lambda\alpha\rho} \gamma_{\epsilon} \Psi_{\rho}, 
\nonumber\\
a_{37} = d_{\lambda}h_{\mu\rho}~{h_{\nu}}^{\rho}~\bar{\Psi}^{\nu} a_{37}^{\epsilon} 
\otimes \Sigma^{\lambda\mu\alpha} \gamma_{\epsilon} \Psi_{\alpha}, \qquad
a_{38} = d_{\lambda}h_{\mu\rho}~{h_{\nu}}^{\rho}~\bar{\Psi}_{\alpha} a_{38}^{\epsilon} 
\otimes \Sigma^{\mu\nu\lambda} \gamma_{\epsilon} \Psi^{\alpha}, 
\nonumber\\
a_{39} = d_{\lambda}h_{\mu\nu}~h^{\mu\nu}~\bar{\Psi}_{\alpha} a_{39}^{\epsilon} 
\otimes \Sigma^{\lambda\alpha\beta} \gamma_{\epsilon} \Psi_{\beta}, \qquad
a_{40} = d_{\lambda}h_{\mu\rho}~{h_{\nu}}^{\rho}~\bar{\Psi}_{\alpha} a_{40}^{\epsilon} 
\otimes \Sigma^{\lambda\nu\alpha} \gamma_{\epsilon} \Psi^{\mu}, 
\nonumber\\
a_{41} = d_{\lambda}h_{\mu\rho}~{h_{\nu}}^{\rho}~\bar{\Psi}_{\alpha} a_{41}^{\epsilon} 
\otimes \Sigma^{\lambda\mu\alpha} \gamma_{\epsilon} \Psi^{\nu}, \qquad
a_{42} = d_{\lambda}h_{\mu\nu}~h~\bar{\Psi}^{\mu} a_{42}^{\epsilon} 
\otimes \Sigma^{\nu\lambda\rho} \gamma_{\epsilon} \Psi_{\rho},
\nonumber\\
a_{43} = d_{\mu}h~h~\bar{\Psi}_{\alpha} a_{43}^{\epsilon} 
\otimes \Sigma^{\mu\alpha\beta} \gamma_{\epsilon} \Psi^{\beta}, \qquad
a_{44} = d_{\lambda}h_{\mu\nu}~h~\bar{\Psi}_{\rho} a_{44}^{\epsilon} 
\otimes \Sigma^{\mu\lambda\rho} \gamma_{\epsilon} \Psi^{\nu},
\nonumber\\
a_{45} = d_{\mu}h~h_{\alpha\beta}~\bar{\Psi}^{\alpha} a_{45}^{\epsilon} 
\otimes \Sigma^{\mu\beta\rho} \gamma_{\epsilon} \Psi^{\rho}, \qquad
a_{46} = d_{\mu}h~h_{\alpha\beta}~\bar{\Psi}_{\rho} a_{46}^{\epsilon} 
\otimes \Sigma^{\mu\alpha\rho} \gamma_{\epsilon} \Psi^{\beta},
\nonumber\\
a_{47} = h_{\mu\lambda}~d_{\nu}u^{\lambda}~\bar{\tilde{\chi}} a_{47}^{\epsilon} 
\otimes \Sigma^{\mu\nu\rho} \gamma_{\epsilon} \Psi_{\rho}, \qquad
a_{48} = h_{\mu\lambda}~d_{\nu}u^{\lambda}\bar{\Psi}_{\rho} a_{48}^{\epsilon} 
\otimes \Sigma^{\mu\nu\rho} \gamma_{\epsilon} \tilde{\chi},
\nonumber\\
a_{49} = d_{\mu}h_{\nu\lambda}~u^{\lambda}~\bar{\tilde{\chi}} a_{49}^{\epsilon} 
\otimes \Sigma^{\mu\nu\rho} \gamma_{\epsilon} \Psi_{\rho}, \qquad
a_{50} = d_{\mu}h_{\nu\lambda}~u^{\lambda}~\bar{\Psi}_{\rho} a_{50}^{\epsilon} 
\otimes \Sigma^{\mu\nu\rho} \gamma_{\epsilon} \tilde{\chi}.
\eea
and
\bea
b_{1}^{\mu} = h^{\mu\nu}~u_{\lambda}~\bar{\Psi}^{\rho} b_{1}^{\epsilon} 
\otimes \Sigma_{\nu\rho\alpha} \gamma_{\epsilon} d^{\lambda}\Psi^{\alpha}, \qquad
b_{2}^{\mu} = h_{\alpha\beta}~u^{\mu}~d^{\alpha}\bar{\Psi}_{\rho} b_{2}^{\epsilon} 
\otimes \Sigma^{\beta\rho\lambda} \gamma_{\epsilon} \Psi_{\lambda},
\nonumber\\
b_{3}^{\mu} = h^{\mu\nu}~u^{\lambda}~\bar{\Psi}^{\rho} b_{3}^{\epsilon} 
\otimes \Sigma_{\nu\rho\lambda} \gamma_{\epsilon} d^{\alpha}\Psi_{\alpha}, \qquad
b_{4}^{\mu} = h_{\alpha\nu}~u_{\lambda}~d^{\alpha}\bar{\Psi}_{\rho} b_{4}^{\epsilon} 
\otimes \Sigma^{\beta\rho\lambda} \gamma_{\epsilon} \Psi^{\mu},
\nonumber\\
b_{5}^{\mu} = h_{\alpha\beta}~u^{\mu}~\bar{\Psi}_{\rho} b_{5}^{\epsilon} 
\otimes \Sigma^{\alpha\rho\lambda} \gamma_{\epsilon} d^{\beta}\Psi_{\lambda}, \qquad
b_{6}^{\mu} = h^{\mu\nu}~u_{\lambda}~d^{\lambda}\bar{\Psi}^{\rho} b_{6}^{\epsilon} 
\otimes \Sigma_{\nu\rho\alpha} \gamma_{\epsilon} \Psi^{\alpha},
\nonumber\\
b_{7}^{\mu} = h_{\alpha\beta}~u_{\lambda}~\bar{\Psi}^{\mu} b_{7}^{\epsilon} 
\otimes \Sigma^{\alpha\lambda\rho} \gamma_{\epsilon} d^{\beta}\Psi_{\rho}, \qquad
b_{8}^{\mu} = h^{\mu\nu}~u^{\lambda}~d^{\alpha}\bar{\Psi}_{\alpha} b_{8}^{\epsilon} 
\otimes \Sigma_{\nu\lambda\rho} \gamma_{\epsilon} \Psi^{\rho},
\nonumber\\
b_{9}^{\mu} = h_{\alpha\beta}~u_{\nu}~d^{\alpha}\bar{\Psi}^{\beta} b_{9}^{\epsilon} 
\otimes \Sigma^{\mu\nu\lambda} \gamma_{\epsilon} \Psi_{\lambda}, \qquad
b_{10}^{\mu} = h^{\mu\nu}~d_{\alpha}u_{\beta}~\bar{\Psi}_{\nu} b_{10}^{\epsilon} 
\otimes \Sigma^{\alpha\beta\lambda} \gamma_{\epsilon} \Psi_{\lambda},
\nonumber\\
b_{11}^{\mu} = h_{\alpha\beta}~u_{\nu}~d^{\alpha}\bar{\Psi}^{\mu} b_{11}^{\epsilon} 
\otimes \Sigma^{\beta\nu\lambda} \gamma_{\epsilon} \Psi_{\lambda}, \qquad
b_{12}^{\mu} = h^{\mu\nu}~d^{\alpha}u^{\beta}~\bar{\Psi}_{\alpha} b_{12}^{\epsilon} 
\otimes \Sigma_{\nu\beta\lambda} \gamma_{\epsilon} \Psi^{\lambda},
\nonumber\\
b_{13}^{\mu} = h_{\alpha\beta}~u_{\nu}~d^{\alpha}\bar{\Psi}^{\nu} b_{13}^{\epsilon} 
\otimes \Sigma^{\mu\beta\lambda} \gamma_{\epsilon} \Psi_{\lambda}, \qquad
b_{14}^{\mu} = h^{\mu\nu}~d^{\alpha}u^{\beta}~\bar{\Psi}_{\beta} b_{14}^{\epsilon} 
\otimes \Sigma_{\nu\alpha\lambda} \gamma_{\epsilon} \Psi^{\lambda},
\nonumber\\
b_{15}^{\mu} = h_{\alpha\beta}~u_{\nu}~d^{\alpha}\bar{\Psi}^{\lambda} b_{15}^{\epsilon} 
\otimes \Sigma^{\mu\nu\beta} \gamma_{\epsilon} \Psi_{\lambda}, \qquad
b_{16} = h^{\mu\nu}~d^{\alpha}u^{\beta}~\bar{\Psi}_{\lambda} b_{16}^{\epsilon} 
\otimes \Sigma_{\nu\alpha\beta} \gamma_{\epsilon} \Psi^{\lambda},
\nonumber\\
b_{17}^{\mu} = h^{\mu\nu}~u_{\lambda}~d_{\nu}\bar{\Psi}_{\rho} b_{17}^{\epsilon} 
\otimes \Sigma^{\lambda\rho\alpha} \gamma_{\epsilon} \Psi_{\alpha}, \qquad
b_{18}^{\mu} = h^{\mu\nu}~d_{\nu}u_{\lambda}~\bar{\Psi}_{\rho} b_{18}^{\epsilon} 
\otimes \Sigma^{\lambda\rho\alpha} \gamma_{\epsilon} \Psi_{\alpha},
\nonumber\\
b_{19}^{\mu} = h_{\alpha\beta}~u^{\beta}~d^{\alpha}\bar{\Psi}_{\rho} b_{19}^{\epsilon} 
\otimes \Sigma^{\mu\rho\lambda} \gamma_{\epsilon} \Psi_{\lambda}, \qquad
b_{20}^{\mu} = h^{\mu\nu}~d_{\alpha}u_{\nu}~\bar{\Psi}_{\rho} b_{20}^{\epsilon} 
\otimes \Sigma^{\alpha\rho\lambda} \gamma_{\epsilon} \Psi_{\lambda},
\nonumber\\
b_{21}^{\mu} = h_{\alpha\beta}~u_{\nu}~d^{\alpha}\bar{\Psi}_{\rho} b_{21}^{\epsilon} 
\otimes \Sigma^{\mu\nu\rho} \gamma_{\epsilon} \Psi^{\beta}, \qquad
b_{22}^{\mu} = h^{\mu\nu}~d^{\alpha}u^{\beta}~\bar{\Psi}^{\rho} b_{22}^{\epsilon} 
\otimes \Sigma_{\rho\alpha\beta} \gamma_{\epsilon} \Psi_{\nu},
\nonumber\\
b_{23}^{\mu} = h^{\mu\nu}~d\cdot u~\bar{\Psi}^{\rho} b_{23}^{\epsilon} 
\otimes \Sigma_{\nu\rho\lambda} \gamma_{\epsilon} \Psi^{\lambda}, \qquad
b_{24}^{\mu} = h^{\mu\nu}~d^{\alpha}u^{\beta}~\bar{\Psi}^{\rho} b_{24}^{\epsilon} 
\otimes \Sigma_{\nu\rho\beta} \gamma_{\epsilon} \Psi_{\alpha},
\nonumber\\
b_{25}^{\mu} = h_{\alpha\beta}~u_{\nu}~d^{\alpha}\bar{\Psi}_{\rho} b_{25}^{\epsilon} 
\otimes \Sigma^{\mu\beta\rho} \gamma_{\epsilon} \Psi^{\nu}, \qquad
b_{26}^{\mu} = h^{\mu\nu}~d^{\alpha}u^{\beta}~\bar{\Psi}^{\rho} b_{26}^{\epsilon} 
\otimes \Sigma_{\nu\rho\alpha} \gamma_{\epsilon} \Psi_{\beta},
\nonumber\\
b_{27}^{\mu} = h_{\alpha\beta}~u_{\nu}~\bar{\Psi}^{\nu} b_{27}^{\epsilon} 
\otimes \Sigma^{\mu\alpha\lambda} \gamma_{\epsilon} d^{\beta}\Psi_{\lambda}, \qquad
b_{28}^{\mu} = h_{\alpha\beta}~u_{\nu}~\bar{\Psi}^{\alpha} b_{28}^{\epsilon} 
\otimes \Sigma^{\mu\nu\lambda} \gamma_{\epsilon} d^{\beta}\Psi_{\lambda},
\nonumber\\
b_{29}^{\mu} = h_{\alpha\beta}~u_{\nu}~\bar{\Psi}^{\lambda} b_{29}^{\epsilon} 
\otimes \Sigma^{\mu\nu\alpha} \gamma_{\epsilon} d^{\beta}\Psi_{\lambda}, \qquad
b_{30}^{\mu} = h_{\alpha\beta}~u^{\beta}~\bar{\Psi}_{\nu} b_{30}^{\epsilon} 
\otimes \Sigma^{\mu\nu\lambda} \gamma_{\epsilon} d^{\alpha}\Psi_{\lambda},
\nonumber\\
b_{31}^{\mu} = h_{\alpha\nu}~u_{\beta}~\bar{\Psi}_{\lambda} b_{31}^{\epsilon} 
\otimes \Sigma^{\mu\nu\lambda} \gamma_{\epsilon} d^{\alpha}\Psi^{\beta}, \qquad
b_{32}^{\mu} = h^{\mu\nu}~u_{\beta}~\bar{\Psi}_{\lambda} b_{32}^{\epsilon} 
\otimes \Sigma^{\lambda\beta\rho} \gamma_{\epsilon} d_{\nu}\Psi_{\rho},
\nonumber\\
b_{33}^{\mu} = h_{\rho\beta}~u_{\nu}~\bar{\Psi}_{\lambda} b_{33}^{\epsilon} 
\otimes \Sigma^{\lambda\nu\rho} \gamma_{\epsilon} d^{\beta}\Psi^{\mu}, \qquad
b_{34}^{\mu} = h_{\alpha\beta}~u_{\nu}~\bar{\Psi}_{\lambda} b_{34}^{\epsilon} 
\otimes \Sigma^{\mu\nu\lambda} \gamma_{\epsilon} d^{\alpha}\Psi_{\beta},
\nonumber\\
b_{35}^{\mu} = h^{\mu\nu}~u^{\lambda}~d^{\alpha}\bar{\Psi}^{\beta} b_{35}^{\epsilon} 
\otimes \Sigma_{\nu\lambda\beta} \gamma_{\epsilon} \Psi_{\alpha}, \qquad
b_{36}^{\mu} = h^{\mu\nu}~u^{\lambda}~\bar{\Psi}_{\alpha} b_{36}^{\epsilon} 
\otimes \Sigma_{\nu\lambda\beta} \gamma_{\epsilon} d^{\alpha}\Psi^{\beta}, 
\nonumber\\
b_{37}^{\mu} = h^{\alpha\beta}~u^{\nu}~d^{\nu}\bar{\Psi}^{\alpha} b_{37}^{\epsilon} 
\otimes \Sigma^{\mu\beta\lambda} \gamma_{\epsilon} \Psi_{\lambda}, \qquad
b_{38}^{\mu} = h_{\alpha\beta}~d_{\nu}u^{\mu}~\bar{\Psi}^{\alpha} b_{38}^{\epsilon} 
\otimes \Sigma^{\nu\beta\lambda} \gamma_{\epsilon} \Psi_{\lambda}, 
\nonumber\\
b_{39}^{\mu} = h^{\alpha\beta}~u_{\nu}~d^{\mu}\bar{\Psi}^{\alpha} b_{39}^{\epsilon} 
\otimes \Sigma^{\nu\beta\lambda} \gamma_{\epsilon} \Psi_{\lambda}, \qquad
b_{40}^{\mu} = h_{\alpha\beta}~d^{\mu}u_{\nu}~\bar{\Psi}^{\alpha} b_{40}^{\epsilon} 
\otimes \Sigma^{\nu\beta\lambda} \gamma_{\epsilon} \Psi_{\lambda}, 
\nonumber\\
b_{41}^{\mu} = h~u_{\nu}~d^{\nu}\bar{\Psi}_{\rho} b_{41}^{\epsilon} 
\otimes \Sigma^{\mu\rho\lambda} \gamma_{\epsilon} \Psi_{\lambda}, \qquad
b_{42}^{\mu} = h~d_{\nu}u^{\mu}~\bar{\Psi}_{\rho} b_{42}^{\epsilon} 
\otimes \Sigma^{\nu\rho\lambda} \gamma_{\epsilon} \Psi_{\lambda},
\nonumber\\
b_{43}^{\mu} = h~u_{\nu}~d^{\mu}\bar{\Psi}_{\rho} b_{43}^{\epsilon} 
\otimes \Sigma^{\nu\rho\lambda} \gamma_{\epsilon} \Psi_{\lambda}, \qquad
b_{44}^{\mu} = h~d^{\mu}u_{\nu}~\bar{\Psi}_{\rho} b_{44}^{\epsilon} 
\otimes \Sigma^{\nu\rho\lambda} \gamma_{\epsilon} \Psi_{\lambda},
\nonumber\\
b_{45}^{\mu} = h_{\alpha\beta}~u_{\nu}~d^{\nu}\bar{\Psi}_{\rho} b_{45}^{\epsilon} 
\otimes \Sigma^{\mu\alpha\rho} \gamma_{\epsilon} \Psi^{\beta}, \qquad
b_{46}^{\mu} = h_{\alpha\beta}~d_{\nu}u^{\mu}~\bar{\Psi}_{\rho} b_{46}^{\epsilon} 
\otimes \Sigma^{\alpha\nu\rho} \gamma_{\epsilon} \Psi^{\beta},
\nonumber\\
b_{47}^{\mu} = h_{\alpha\beta}~u_{\nu}~d^{\mu}\bar{\Psi}_{\rho} b_{47}^{\epsilon} 
\otimes \Sigma^{\alpha\nu\rho} \gamma_{\epsilon} \Psi^{\beta}, \qquad
b_{48}^{\mu} = h_{\alpha\beta}~d^{\mu}u_{\nu}~\bar{\Psi}_{\rho} b_{48}^{\epsilon} 
\otimes \Sigma^{\alpha\nu\rho} \gamma_{\epsilon} \Psi^{\beta},
\nonumber\\
b_{49}^{\mu} = h_{\nu\lambda}~u^{\lambda}~d^{\mu}\bar{\Psi}_{\alpha} b_{49}^{\epsilon} 
\otimes \Sigma^{\nu\alpha\beta} \gamma_{\epsilon} \Psi_{\beta}, \qquad
b_{50}^{\mu} = h_{\nu\lambda}~d^{\mu}u^{\lambda}~\bar{\Psi}_{\alpha} b_{50}^{\epsilon} 
\otimes \Sigma^{\nu\alpha\beta} \gamma_{\epsilon} \Psi_{\beta}
\nonumber\\
b_{51}^{\mu} = h_{\nu\lambda}~d^{\lambda}u^{\mu}~\bar{\Psi}_{\alpha} b_{51}^{\epsilon} 
\otimes \Sigma^{\nu\alpha\beta} \gamma_{\epsilon} \Psi_{\beta}, \qquad
b_{52}^{\mu} = h~u_{\nu}~d_{\lambda}\bar{\Psi}^{\lambda} b_{52}^{\epsilon} 
\otimes \Sigma^{\mu\nu\rho} \gamma_{\epsilon} \Psi_{\rho},
\nonumber\\
b_{53}^{\mu} = h~d_{\alpha}u_{\beta}~\bar{\Psi}^{\mu} b_{53}^{\epsilon} 
\otimes \Sigma^{\alpha\beta\rho} \gamma_{\epsilon} \Psi_{\rho}, \qquad
b_{54}^{\mu} = h_{\alpha\lambda}~d^{\lambda}u_{\beta}~\bar{\Psi}^{\mu} b_{54}^{\epsilon} 
\otimes \Sigma^{\alpha\beta\rho} \gamma_{\epsilon} \Psi_{\rho}
\nonumber\\
b_{55}^{\mu} = h_{\alpha\lambda}~d^{\lambda}u_{\beta}~\bar{\Psi}_{\nu} b_{55}^{\epsilon} 
\otimes \Sigma^{\alpha\beta\nu} \gamma_{\epsilon} \Psi^{\mu}, \qquad
b_{56}^{\mu} = h~u_{\nu}~\bar{\Psi}_{\rho} b_{56}^{\epsilon} 
\otimes \Sigma^{\nu\rho\lambda} \gamma_{\epsilon} d^{\mu}\Psi_{\lambda},
\nonumber\\
b_{57}^{\mu} = h_{\alpha\beta}~u_{\lambda}~\bar{\Psi}_{\rho} b_{57}^{\epsilon} 
\otimes \Sigma^{\alpha\lambda\rho} \gamma_{\epsilon} d^{\mu}\Psi^{\beta}, \qquad
b_{58}^{\mu} = h_{\nu\lambda}~u^{\lambda}~\bar{\Psi}_{\alpha} b_{58}^{\epsilon} 
\otimes \Sigma^{\nu\alpha\beta} \gamma_{\epsilon} d^{\mu}\Psi_{\beta},
\nonumber\\
b_{59}^{\mu} = h_{\alpha\beta}~u_{\nu}~\bar{\Psi}^{\alpha} b_{59}^{\epsilon} 
\otimes \Sigma^{\nu\beta\lambda} \gamma_{\epsilon} d^{\mu}\Psi_{\lambda}, \qquad
b_{60}^{\mu} = h_{\alpha\beta}~u_{\nu}~\bar{\Psi}^{\alpha} b_{60}^{\epsilon} 
\otimes \Sigma^{\mu\beta\rho} \gamma_{\epsilon} d^{\nu}\Psi_{\rho},
\nonumber\\
b_{61}^{\mu} = h~u_{\nu}~d_{\lambda}\bar{\Psi}_{\rho} b_{61}^{\epsilon} 
\otimes \Sigma^{\mu\nu\rho} \gamma_{\epsilon} \Psi^{\lambda}, \qquad
b_{62}^{\mu} = h~u_{\nu}~\bar{\Psi}^{\lambda} b_{62}^{\epsilon} 
\otimes \Sigma^{\mu\nu\rho} \gamma_{\epsilon} d_{\lambda}\Psi_{\rho},
\nonumber\\
b_{63}^{\mu} = h~u_{\nu}~\bar{\Psi}_{\rho} b_{63}^{\epsilon} 
\otimes \Sigma^{\mu\rho\lambda} \gamma_{\epsilon} d^{\nu}\Psi_{\lambda}, \qquad
b_{64}^{\mu} = h_{\alpha\beta}~u_{\nu}~\bar{\Psi}_{\rho} b_{64}^{\epsilon} 
\otimes \Sigma^{\mu\alpha\rho} \gamma_{\epsilon} d^{\nu}\Psi^{\beta},
\nonumber\\
b_{65}^{\mu} = h~u_{\nu}~\bar{\Psi}_{\rho} b_{65}^{\epsilon} 
\otimes \Sigma^{\mu\nu\rho} \gamma_{\epsilon} d^{\lambda}\Psi_{\lambda}, \qquad
b_{66} = h~d_{\alpha}u_{\beta}~\bar{\Psi}_{\rho} b_{66}^{\epsilon} 
\otimes \Sigma^{\alpha\beta\rho} \gamma_{\epsilon} \Psi^{\mu},
\nonumber\\
b_{67}^{\mu} = d^{\lambda}h^{\mu\nu}~u_{\lambda}~\bar{\Psi}^{\alpha} b_{67}^{\epsilon} 
\otimes \Sigma_{\nu\alpha\beta} \gamma_{\epsilon} \Psi^{\beta}, \qquad
b_{68}^{\mu} = d^{\alpha}h^{\mu\nu}~u^{\beta}~\bar{\Psi}^{\rho} b_{68}^{\epsilon} 
\otimes \Sigma_{\nu\rho\beta} \gamma_{\epsilon} \Psi_{\alpha},
\nonumber\\
b_{69}^{\mu} = d^{\alpha}h^{\mu\nu}~u^{\beta}~\bar{\Psi}_{\alpha} b_{69}^{\epsilon} 
\otimes \Sigma_{\nu\beta\rho} \gamma_{\epsilon} \Psi^{\rho}, \qquad
b_{70}^{\mu} = d^{\alpha}h^{\mu\nu}~u^{\beta}~\bar{\Psi}_{\nu} b_{70}^{\epsilon} 
\otimes \Sigma_{\alpha\beta\lambda} \gamma_{\epsilon} \Psi^{\lambda},
\nonumber\\
b_{71}^{\mu} = d^{\alpha}h^{\mu\nu}~u^{\beta}~\bar{\Psi}_{\beta} b_{71}^{\epsilon} 
\otimes \Sigma_{\nu\alpha\lambda} \gamma_{\epsilon} \Psi^{\lambda}, \qquad
b_{72}^{\mu} = d^{\alpha}h^{\mu\nu}~u^{\beta}~\bar{\Psi}^{\lambda} b_{72}^{\epsilon} 
\otimes \Sigma_{\nu\alpha\beta} \gamma_{\epsilon} \Psi_{\lambda},
\nonumber\\
b_{73}^{\mu} = d_{\nu}h^{\mu\lambda}~u_{\lambda}~\bar{\Psi}_{\alpha} b_{73}^{\epsilon} 
\otimes \Sigma^{\nu\alpha\beta} \gamma_{\epsilon} \Psi_{\beta}, \qquad
b_{74}^{\mu} = d_{\alpha}h^{\mu\nu}~u_{\beta}~\bar{\Psi}_{\lambda} b_{74}^{\epsilon} 
\otimes \Sigma^{\alpha\beta\lambda} \gamma_{\epsilon} \Psi_{\nu},
\nonumber\\
b_{75}^{\mu} = d^{\alpha}h^{\mu\nu}~u^{\beta}~\bar{\Psi}^{\lambda} b_{75}^{\epsilon} 
\otimes \Sigma_{\nu\alpha\lambda} \gamma_{\epsilon} \Psi_{\beta}, \qquad
b_{76}^{\mu} = d^{\lambda}h_{\lambda\alpha}~u_{\beta}~\bar{\Psi}_{\rho} b_{76}^{\epsilon} 
\otimes \Sigma^{\alpha\beta\rho} \gamma_{\epsilon} \Psi^{\mu},
\nonumber\\
b_{77}^{\mu} = d^{\lambda}h_{\lambda\alpha}~u_{\beta}~\bar{\Psi}^{\mu} b_{77}^{\epsilon} 
\otimes \Sigma^{\alpha\beta\rho} \gamma_{\epsilon} \Psi_{\rho}, \qquad
b_{78}^{\mu} = d_{\nu}h_{\alpha\beta}~u^{\mu}~\bar{\Psi}^{\alpha} b_{78}^{\epsilon} 
\otimes \Sigma^{\nu\beta\lambda} \gamma_{\epsilon} \Psi_{\lambda},
\nonumber\\
b_{79}^{\mu} = d^{\mu}h_{\alpha\beta}~u_{\nu}~\bar{\Psi}^{\alpha} b_{79}^{\epsilon} 
\otimes \Sigma^{\nu\beta\lambda} \gamma_{\epsilon} \Psi_{\lambda}, \qquad
b_{80}^{\mu} = d_{\nu}h~u^{\mu}~\bar{\Psi}_{\alpha} b_{80}^{\epsilon} 
\otimes \Sigma^{\nu\alpha\beta} \gamma_{\epsilon} \Psi_{\beta},
\nonumber\\
b_{81}^{\mu} = d^{\mu}h~u_{\nu}~\bar{\Psi}_{\alpha} b_{81}^{\epsilon} 
\otimes \Sigma^{\nu\alpha\beta} \gamma_{\epsilon} \Psi_{\beta}, \qquad
b_{82}^{\mu} = d_{\nu}h_{\alpha\beta}~u^{\mu}~\bar{\Psi}_{\rho} b_{82}^{\epsilon} 
\otimes \Sigma^{\nu\alpha\rho} \gamma_{\epsilon} \Psi^{\beta},
\nonumber\\
b_{83}^{\mu} = d^{\mu}h_{\alpha\beta}~u_{\nu}~\bar{\Psi}_{\lambda} b_{83}^{\epsilon} 
\otimes \Sigma^{\nu\alpha\lambda} \gamma_{\epsilon} \Psi^{\beta}, \qquad
b_{84}^{\mu} = d^{\mu}h_{\nu\lambda}~u^{\lambda}~\bar{\Psi}_{\alpha} b_{84}^{\epsilon} 
\otimes \Sigma^{\nu\alpha\beta} \gamma_{\epsilon} \Psi^{\beta},
\nonumber\\
b_{85}^{\mu} = d_{\alpha}h~u_{\beta}~\bar{\Psi}^{\mu} b_{85}^{\epsilon} 
\otimes \Sigma^{\alpha\beta\rho} \gamma_{\epsilon} \Psi_{\rho}, \qquad
b_{86}^{\mu} = d_{\alpha}h~u_{\beta}~\bar{\Psi}_{\rho} b_{86}^{\epsilon} 
\otimes \Sigma^{\alpha\beta\rho} \gamma_{\epsilon} \Psi^{\mu}, 
\nonumber\\
b_{87}^{\mu} = d_{\beta}h^{\alpha\beta}~u^{\nu}~\bar{\Psi}^{\alpha} b_{87}^{\epsilon} 
\otimes \Sigma^{\mu\nu\lambda} \gamma_{\epsilon} \Psi_{\lambda}, \qquad
b_{88}^{\mu} = d^{\beta}h_{\alpha\beta}~u_{\nu}~\bar{\Psi}^{\nu} b_{88}^{\epsilon} 
\otimes \Sigma^{\mu\alpha\lambda} \gamma_{\epsilon} \Psi_{\lambda}, 
\nonumber\\
b_{89}^{\mu} = d^{\beta}h_{\alpha\beta}~u_{\nu}~\bar{\Psi}_{\lambda} b_{89}^{\epsilon} 
\otimes \Sigma^{\mu\nu\alpha} \gamma_{\epsilon} \Psi^{\lambda}, \qquad
b_{90}^{\mu} = d_{\beta}h^{\alpha\beta}~u_{\nu}~\bar{\Psi}_{\lambda} b_{90}^{\epsilon} 
\otimes \Sigma^{\mu\nu\lambda} \gamma_{\epsilon} \Psi_{\alpha}, 
\nonumber\\
b_{91}^{\mu} = d^{\beta}h_{\alpha\beta}~u_{\nu}~\bar{\Psi}_{\lambda} b_{91}^{\epsilon} 
\otimes \Sigma^{\mu\alpha\lambda} \gamma_{\epsilon} \Psi^{\nu}, \qquad
b_{92}^{\mu} = d_{\lambda}h_{\alpha\beta}~u^{\lambda}~\bar{\Psi}^{\alpha} b_{92}^{\epsilon} 
\otimes \Sigma^{\mu\beta\rho} \gamma_{\epsilon} \Psi_{\rho},
\nonumber\\
b_{93}^{\mu} = d^{\nu}h~u_{\nu}~\bar{\Psi}_{\alpha} b_{93}^{\epsilon} 
\otimes \Sigma^{\mu\alpha\beta} \gamma_{\epsilon} \Psi_{\beta}, \qquad
b_{94}^{\mu} = d_{\lambda}h~u_{\nu}~\bar{\Psi}^{\lambda} b_{94}^{\epsilon} 
\otimes \Sigma^{\mu\nu\rho} \gamma_{\epsilon} \Psi_{\rho},
\nonumber\\
b_{95}^{\mu} = d_{\lambda}h_{\alpha\beta}~u^{\lambda}~\bar{\Psi}_{\rho} b_{95}^{\epsilon} 
\otimes \Sigma^{\mu\alpha\rho} \gamma_{\epsilon} \Psi^{\beta}, \qquad
b_{96}^{\mu} = d_{\lambda}h~u_{\nu}~\bar{\Psi}_{\rho} b_{96}^{\epsilon} 
\otimes \Sigma^{\mu\nu\rho} \gamma_{\epsilon} \Psi^{\lambda},
\nonumber\\
b_{97}^{\mu} = h_{\alpha\beta}~d^{\alpha}u_{\lambda}~\bar{\Psi}^{\beta} b_{97}^{\epsilon} 
\otimes \Sigma^{\mu\lambda\rho} \gamma_{\epsilon} \Psi_{\rho}, \qquad
b_{98}^{\mu} = h_{\alpha\beta}~d^{\alpha}u_{\lambda}~\bar{\Psi}^{\lambda} b_{98}^{\epsilon} 
\otimes \Sigma^{\mu\beta\rho} \gamma_{\epsilon} \Psi_{\rho},
\nonumber\\
b_{99}^{\mu} = h_{\alpha\beta}~d^{\alpha}u_{\lambda}~\bar{\Psi}_{\rho} b_{99}^{\epsilon} 
\otimes \Sigma^{\mu\beta\lambda} \gamma_{\epsilon} \Psi^{\rho}, \qquad
b_{100}^{\mu} = h^{\alpha\beta}~d_{\alpha}u_{\beta}~\bar{\Psi}_{\rho} b_{100}^{\epsilon} 
\otimes \Sigma^{\mu\rho\lambda} \gamma_{\epsilon} \Psi_{\lambda}
\nonumber\\
b_{101}^{\mu} = h_{\alpha\beta}~d^{\alpha}u_{\lambda}~\bar{\Psi}_{\rho} b_{101}^{\epsilon} 
\otimes \Sigma^{\mu\lambda\rho} \gamma_{\epsilon} \Psi_{\beta}, \qquad
b_{102}^{\mu} = h_{\alpha\beta}~d^{\alpha}u_{\lambda}~\bar{\Psi}_{\rho} b_{102}^{\epsilon} 
\otimes \Sigma^{\mu\beta\rho} \gamma_{\epsilon} \Psi^{\lambda},
\nonumber\\
b_{103}^{\mu} = h_{\alpha\beta}~d\cdot u~\bar{\Psi}^{\alpha} b_{103}^{\epsilon} 
\otimes \Sigma^{\mu\beta\rho} \gamma_{\epsilon} \Psi_{\rho}, \qquad
b_{104}^{\mu} = h~d\cdot u~\bar{\Psi}_{\alpha} b_{104}^{\epsilon} 
\otimes \Sigma^{\mu\alpha\rho} \gamma_{\epsilon} \Psi_{\beta}
\nonumber\\
b_{105}^{\mu} = h_{\alpha\beta}~d\cdot u~\bar{\Psi}_{\rho} b_{105}^{\epsilon} 
\otimes \Sigma^{\mu\alpha\rho} \gamma_{\epsilon} \Psi^{\beta}, \qquad
b_{106}^{\mu} = h~d_{\alpha}u_{\beta}~\bar{\Psi}^{\alpha} b_{106}^{\epsilon} 
\otimes \Sigma^{\mu\beta\rho} \gamma_{\epsilon} \Psi_{\rho},
\nonumber\\
b_{107}^{\mu} = h~d_{\alpha}u_{\beta}~\bar{\Psi}_{\rho} b_{107}^{\epsilon} 
\otimes \Sigma^{\mu\beta\rho} \gamma_{\epsilon} \Psi^{\alpha}, \qquad
b_{108}^{\mu} = h_{\alpha\beta}~d_{\lambda}u^{\alpha}~\bar{\Psi}^{\mu} b_{108}^{\epsilon} 
\otimes \Sigma^{\beta\lambda\rho} \gamma_{\epsilon} \Psi_{\rho}
\nonumber\\
b_{109}^{\mu} = d_{\alpha}h_{\lambda\beta}~u^{\lambda}~\bar{\Psi}^{\alpha} b_{109}^{\epsilon} 
\otimes \Sigma^{\mu\beta\rho} \gamma_{\epsilon} \Psi_{\rho}, \qquad
b_{110}^{\mu} = h_{\alpha\beta}~d_{\lambda}u^{\alpha}~\bar{\Psi}^{\beta} b_{110}^{\epsilon} 
\otimes \Sigma^{\mu\lambda\rho} \gamma_{\epsilon} \Psi_{\rho}, 
\nonumber\\
b_{111}^{\mu} = d_{\alpha}h_{\lambda\beta}~u^{\lambda}~\bar{\Psi}^{\beta} b_{111}^{\epsilon} 
\otimes \Sigma^{\mu\alpha\rho} \gamma_{\epsilon} \Psi_{\rho}, \qquad
b_{112}^{\mu} = h_{\alpha\beta}~d_{\lambda}u^{\alpha}~\bar{\Psi}^{\lambda} b_{112}^{\epsilon} 
\otimes \Sigma^{\mu\beta\rho} \gamma_{\epsilon} \Psi_{\rho},
\nonumber\\
b_{113}^{\mu} = d_{\alpha}h_{\lambda\beta}~u^{\lambda}~\bar{\Psi}^{\mu} b_{113}^{\epsilon} 
\otimes \Sigma^{\alpha\beta\rho} \gamma_{\epsilon} \Psi_{\rho}, \qquad
b_{114}^{\mu} = h_{\alpha\beta}~d_{\lambda}u^{\alpha}~\bar{\Psi}_{\rho} b_{114}^{\epsilon} 
\otimes \Sigma^{\mu\beta\lambda} \gamma_{\epsilon} \Psi^{\rho},
\nonumber\\
b_{115}^{\mu} = d_{\alpha}h_{\lambda\beta}~u^{\lambda}~\bar{\Psi}_{\rho} b_{115}^{\epsilon} 
\otimes \Sigma^{\mu\alpha\beta} \gamma_{\epsilon} \Psi^{\rho}, \qquad
b_{116}^{\mu} = h_{\alpha\beta}~d_{\lambda}u^{\alpha}~\bar{\Psi}_{\rho} b_{116}^{\epsilon} 
\otimes \Sigma^{\mu\lambda\rho} \gamma_{\epsilon} \Psi^{\beta},
\nonumber\\
b_{117}^{\mu} = d_{\alpha}h_{\lambda\beta}~u^{\lambda}~\bar{\Psi}_{\rho} b_{117}^{\epsilon} 
\otimes \Sigma^{\mu\alpha\rho} \gamma_{\epsilon} \Psi^{\beta}, \qquad
b_{118}^{\mu} = h_{\alpha\beta}~d_{\lambda}u^{\alpha}~\bar{\Psi}_{\rho} b_{118}^{\epsilon} 
\otimes \Sigma^{\mu\beta\rho} \gamma_{\epsilon} \Psi^{\lambda},
\nonumber\\
b_{119}^{\mu} = d_{\alpha}h_{\lambda\beta}~u^{\lambda}~\bar{\Psi}_{\rho} b_{119}^{\epsilon} 
\otimes \Sigma^{\alpha\beta\rho} \gamma_{\epsilon} \Psi^{\mu}, \qquad
b_{120}^{\mu} = h~d_{\alpha}u_{\beta}~\bar{\Psi}^{\beta} b_{120}^{\epsilon} 
\otimes \Sigma^{\mu\alpha\rho} \gamma_{\epsilon} \Psi_{\rho}
\nonumber\\
b_{121}^{\mu} = d_{\alpha}h~u_{\nu}~\bar{\Psi}^{\nu} b_{121}^{\epsilon} 
\otimes \Sigma^{\mu\alpha\rho} \gamma_{\epsilon} \Psi_{\rho}, \qquad
b_{122}^{\mu} = h~d_{\alpha}u_{\beta}~\bar{\Psi}_{\rho} b_{122}^{\epsilon} 
\otimes \Sigma^{\mu\alpha\rho} \gamma_{\epsilon} \Psi^{\beta},
\nonumber\\
b_{123}^{\mu} = d_{\alpha}h~u_{\nu}~\bar{\Psi}_{\rho} b_{123}^{\epsilon} 
\otimes \Sigma^{\mu\alpha\rho} \gamma_{\epsilon} \Psi^{\nu}, \qquad
b_{124}^{\mu} = h_{\alpha\beta}~d_{\lambda}u^{\alpha}~\bar{\Psi}_{\rho} b_{124}^{\epsilon} 
\otimes \Sigma^{\beta\lambda\rho} \gamma_{\epsilon} \Psi^{\mu}
\nonumber\\
b_{125}^{\mu} = d_{\alpha}h_{\lambda\beta}~u^{\lambda}~\bar{\Psi}_{\rho} b_{125}^{\epsilon} 
\otimes \Sigma^{\mu\beta\rho} \gamma_{\epsilon} \Psi^{\alpha}, \qquad
b_{126}^{\mu} = h_{\nu\lambda}~u^{\lambda}~d_{\alpha}\bar{\Psi}^{\alpha} b_{126}^{\epsilon} 
\otimes \Sigma^{\mu\nu\rho} \gamma_{\epsilon} \Psi_{\rho},
\nonumber\\
b_{127}^{\mu} = h_{\nu\lambda}~u^{\lambda}~d_{\alpha}\bar{\Psi}_{\rho} b_{127}^{\epsilon} 
\otimes \Sigma^{\mu\nu\rho} \gamma_{\epsilon} \Psi^{\alpha}, \qquad
b_{128}^{\mu} = h_{\nu\lambda}~u^{\lambda}~\bar{\Psi}_{\rho} b_{128}^{\epsilon} 
\otimes \Sigma^{\mu\nu\rho} \gamma_{\epsilon} d_{\alpha}\Psi^{\alpha}
\nonumber\\
b_{129}^{\mu} = h_{\nu\lambda}~u^{\lambda}~\bar{\Psi}^{\alpha} b_{129}^{\epsilon} 
\otimes \Sigma^{\mu\nu\rho} \gamma_{\epsilon} d_{\alpha}\Psi_{\rho}.
\eea
Now we compute
\bea
A_{hu} - sb \equiv A_{hu} - (d_{Q}b - i d_{\mu}b^{\mu}) =
\nonumber\\
i~\Bigl\{ h_{\mu\lambda}~d^{\lambda}u_{\nu}~d^{\mu}\bar{\Psi}_{\alpha}~
\Bigl[ 4(t^{\epsilon})^{2} + {1\over 2} a_{4}^{\epsilon} + b_{17}^{\epsilon} + b_{18}^{\epsilon}\Bigl] \otimes 
\Sigma^{\nu\alpha\beta}\gamma_{\epsilon}\Psi_{\beta}
\nonumber\\
+ h_{\mu\lambda}~d_{\nu}u^{\lambda}~d^{\mu}\bar{\Psi}_{\alpha}~
\Bigl[ - 4(t^{\epsilon})^{2} + {1\over 2} a_{4}^{\epsilon} + b_{19}^{\epsilon} + b_{20}^{\epsilon}\Bigl] \otimes 
\Sigma^{\nu\alpha\beta}\gamma_{\epsilon}\Psi_{\beta}
\nonumber\\
+ u^{\lambda}~d_{\mu}h_{\lambda\nu}~d^{\nu}\bar{\Psi}_{\alpha}~(b_{19}^{\epsilon} + b_{73}^{\epsilon})
\otimes \Sigma^{\mu\alpha\beta}\gamma_{\epsilon}\Psi_{\beta}
\nonumber\\
+ d_{\mu}u_{\nu}~d_{\lambda}h^{\mu\nu}~\bar{\Psi}_{\alpha}~(a_{39}^{\epsilon} + b_{73}^{\epsilon} + b_{100}^{\epsilon}) 
\otimes \Sigma^{\lambda\alpha\beta}\gamma_{\epsilon}\Psi_{\beta}
\nonumber\\
+ d_{\mu}d_{\nu}u_{\lambda}~h^{\mu\nu}~\bar{\Psi}_{\alpha}~\Bigl( {1\over 2} a_{33}^{\epsilon} + b_{18}^{\epsilon}\Bigl) \otimes 
\Sigma^{\lambda\alpha\beta}\gamma_{\epsilon}\Psi_{\beta}
\nonumber\\
+ d_{\mu}d_{\lambda}u_{\nu}~h^{\mu\nu}~\bar{\Psi}_{\alpha}~
\Bigl( {1\over 2} a_{33}^{\epsilon} + a_{39}^{\epsilon} + b_{20}^{\epsilon} + b_{100}^{\epsilon}\Bigl) \otimes 
\Sigma^{\lambda\alpha\beta}\gamma_{\epsilon}\Psi_{\beta} \Bigl\} + \cdots
\eea
where 
$\cdots$
are other terms. If we equate to $0$ this expression we find out 
$
(t^{\epsilon})^{2} = 0 \quad \Rightarrow \quad t^{\epsilon} = 0
$
which proves that there are no solutions.
$\qed$

\section{Conclusions}
The negative result for the interaction between spins $3/2$ and $2$ hints that there might be problems for the 
interaction of other higher spins also.

We remark that our consideration is {\bf not} based on supersymmetry. To have a supersymmetric model one would have 
to include the supersymmetric partners of the ghost fields and this could lead to no-go results as in \cite{no-susy}.


\end{document}